\documentclass[sigconf]{acmart}

\usepackage{enumitem}

\usepackage{amssymb}
\usepackage[ruled]{algorithm2e}
\usepackage{url}
\usepackage{multirow}
\usepackage{subfigure}
\usepackage{bm}
\usepackage{verbatim}
\usepackage{ragged2e}
\usepackage{caption}
\usepackage{subcaption}
\usepackage{graphicx}
\usepackage[normalem]{ulem}
\useunder{\uline}{\ul}{}
\usepackage{amsmath}
\usepackage{longtable}
\usepackage{adjustbox}
\usepackage{microtype}
\usepackage{setspace}
\usepackage{hyperref}
\usepackage{float}
\usepackage{balance}
\newcommand{\ie}{\emph{i.e., }}

\newcommand{\aka}

 \newcommand{\bym}[1]{{#1}}

\AtBeginDocument{%
  }

\setcopyright{acmlicensed}
\copyrightyear{2018}
\acmYear{2018}
\acmDOI{XXXXXXX.XXXXXXX}

\acmConference[Conference acronym 'XX]{Make sure to enter the correct
  conference title from your rights confirmation emai}{June 03--05,
  2018}{Woodstock, NY}
\acmISBN{978-1-4503-XXXX-X/18/06}

\usepackage{colortbl}  
\usepackage{xcolor}
\usepackage{url}
\usepackage{multirow}
\usepackage{subfigure}
\usepackage{arydshln}
\usepackage{tabularx}
\usepackage{enumitem}
\usepackage{array}   

\begin{document}


\title{Causality-Enhanced Behavior Sequence Modeling in LLMs for Personalized Recommendation}


\author{Yang Zhang}
\authornote{Equal contribution.}
\orcid{0000-0002-7863-5183}
\affiliation{
  \institution{National University of Singapore}
  \city{Kent Ridge}
  \country{Singapore}
}
\email{zyang1580@gmail.com}

\author{Juntao You}
\authornotemark[1]
\affiliation{
  \institution{University of Science and Technology of China}
  \city{Heifei}
  \country{China}
}
\email{ustcyjt@mail.ustc.edu.cn}

\author{Yimeng Bai}
\affiliation{
  \institution{University of Science and Technology of China}
  \city{Heifei}
  \country{China}
}
\email{baiyimeng77@mail.ustc.edu.cn}

\author{Jizhi Zhang}
\affiliation{
  \institution{University of Science and Technology of China}
  \city{Heifei}
  \country{China}
}
\email{cdzhangjizhi@mail.ustc.edu.cn}

\author{Keqin Bao}
\affiliation{
  \institution{University of Science and Technology of China}
  \city{Heifei}
  \country{China}
}
\email{baokq@mail.ustc.edu.cn}

\author{Wenjie Wang}
\affiliation{
  \institution{National University of Singapore}
  \city{Kent Ridge}
  \country{Singapore}
}
\email{wenjiewang96@gmail.com}

\author{Tat-Seng Chua 
}
\affiliation{
  \institution{National University of Singapore}
  \city{Kent Ridge}
  \country{Singapore}
}
\email{chuats@comp.nus.edu.sg}

\renewcommand{\shortauthors}{Yang Zhang et al.}

\begin{abstract}

Recent advancements in recommender systems have focused on leveraging Large Language Models (LLMs) to improve user preference modeling, yielding promising outcomes.
However, current LLM-based approaches struggle to fully leverage user behavior sequences, resulting in suboptimal preference modeling for personalized recommendations. In this study, we propose a novel \textit{Counterfactual Fine-Tuning} (CFT) method to address this issue by explicitly emphasizing the role of behavior sequences when generating recommendations. Specifically, we employ counterfactual reasoning to identify the causal effects of behavior sequences on model output and introduce a task that directly fits the ground-truth labels based on these effects, achieving the goal of explicit emphasis. Additionally, we develop a token-level weighting mechanism to adjust the emphasis strength for different item tokens, reflecting the diminishing influence of behavior sequences from earlier to later tokens during predicting an item. Extensive experiments on real-world datasets demonstrate that CFT effectively improves behavior sequence modeling. Our codes are available at 
\url{https://github.com/itsmeyjt/CFT}.
\end{abstract}






\maketitle

\section{Introduction}

In recent years, there has been growing enthusiasm for developing recommender systems based on Large Language Models (LLMs)~\cite{zoranker,llm4trdrec,prompt_distill, agentcf, lcrec,binllm,tallrec,llmrec_survey}, with the expectation that LLMs' advanced capabilities could bring a new revolution to recommendation fields. However, the core of recommendation lies in user preference modeling~\cite{user_modeling}, which inherently differs from the language processing tasks LLMs were originally designed for~\cite{tallrec}. To bridge this gap, current approaches often construct instruction data from historical user behavior sequences and then use it to fine-tune LLMs~\cite{instructrec,binllm,lcrec}. This process equips LLMs with the ability to model user behavior, allowing them to infer user preferences based on the provided behavior sequences and their internalized knowledge. Till now, numerous efforts have been made in this direction, leading to significant progress~\cite{lcrec,recgpt,d3,transrec,bigrec}.

\begin{figure}[t]
    \centering
    \subfigure[\textbf{Amazon Books}]{\includegraphics[width=0.23\textwidth]{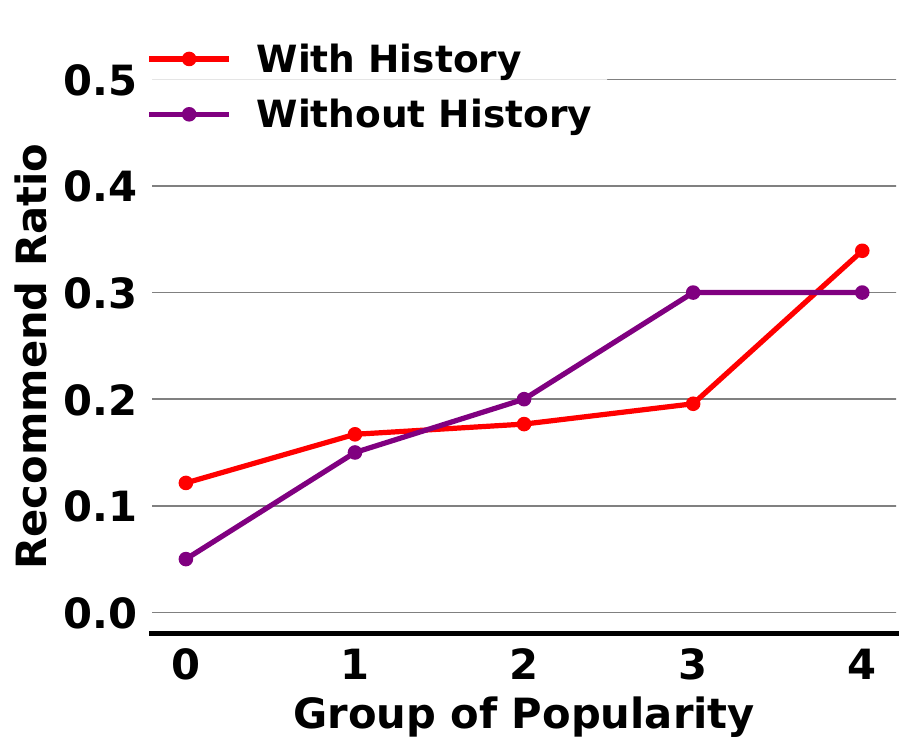}}
    \subfigure[ \textbf{Amazon Games}]{ \includegraphics[width=0.23\textwidth]{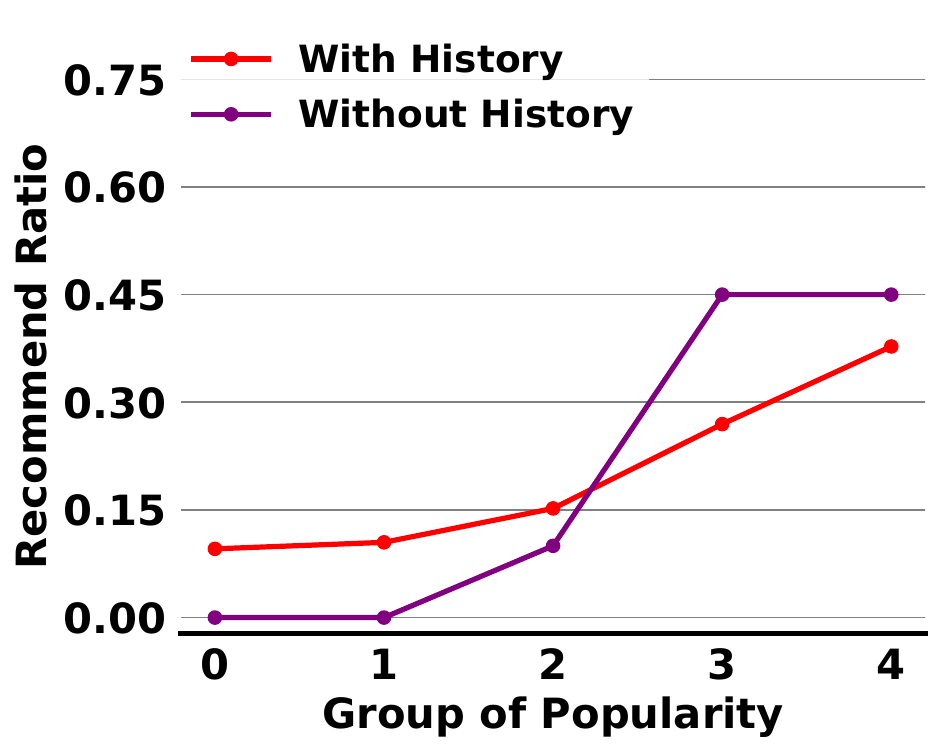}}
    \caption{Recommendation distribution comparison for LLM-based method BIGRec~\cite{bigrec} between with and without inputting historical behavior sequences on Amazon data~\cite{amazon}.
    The result reveals that whether or not behavior sequences are input leads to similar distributions of recommended items, indicating that the information is not fully utilized.
    }
    \label{fig:cmp-his}
    \Description{..}
\end{figure}

Despite the remarkable progress, we argue that current fine-tuning methods may not fully harness the potential of LLMs in effectively utilizing the behavior sequences. When building the recommender models with LLMs, 
information unrelated to the input behavior sequence—such as pre-training knowledge (world knowledge) triggered by the task instruction\footnote{LLMs can  provide recommendations with only task instructions, \textit{e.g.,} ChatGPT can directly respond to the prompt, ``please recommend a movie to me."}—can also be utilized for prediction.
This information is simpler to utilize than complex behavior-related information. 
Consequently, LLMs may tend to over-rely on this information when fitting prediction targets, resulting in insufficient utilization of behavior sequences. One of our findings indicates this potential insufficient utilization: when removing the user behavior sequence from the input, the LLM-based recommender still produced similar recommendation distributions to those generated with the sequence included, as shown in Figure~\ref{fig:cmp-his}.

\begin{figure}[t]
    \centering
    \includegraphics[width=2.2in,height=1.475in]{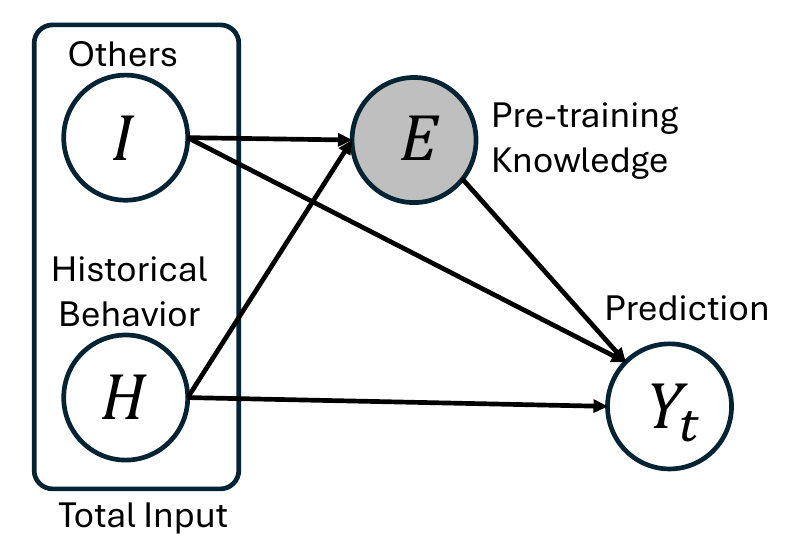}
    \caption{
    Causal graph illustrating the prediction generation process in LLM-based recommendation: the input behavior sequences \(H\) and other input information \(I\) (e.g., task instructions) can influence the ($t$-th token) prediction $Y_t$ directly or indirectly by activating the pre-training knowledge \(E\). 
    }
    \label{fig:causal_graph}
    \Description{..}
\end{figure}

Given the central role of behavior modeling in recommendations, it is crucial to address the problem of insufficient utilization of behavior sequences in LLM-based recommendations. To tackle this issue, we first employ \textit{causal language}~\cite{pearl2009causality} for a qualitative analysis. As illustrated in Figure~\ref{fig:causal_graph}, given the inputs, LLMs can generate predictions through various causal paths, some of which do not originate from the behavior sequence. The insufficient utilization of behavior sequences arises from the tendency to overlook the effects of behavior sequence-originated paths compared to the others. Therefore, to enhance behavior sequence modeling, 
the key is to identify the effects of these behavior sequence-originated paths, which correspond to the causal effect of behavior sequences on the predictions, and emphasize these effects during the tuning process.

To this end, we propose a novel fine-tuning method, which explicitly emphasizes the effects of behavior sequences on LLM predictions during tuning to enhance behavior modeling. Specifically, we leverage counterfactual reasoning~\cite{pearl2009causality,pearl2016primer} to estimate the effects as the difference between normal predictions (i.e., those generated using behavior sequences) and counterfactual predictions (i.e., those generated without them).  
We then introduce a new task that directly uses the estimated effects to fit the ground-truth labels, explicitly attributing the label occurrences to these effects, achieving the intended emphasis.  
This task is integrated into the tuning in a multi-task manner~\cite{multitask}, allowing us to retain the original task of fitting labels with normal predictions, thereby preserving other valuable information relevant to the task.  

Taking a further step, we observe that in the recommendation task, each sample's prediction usually involves forecasting a sequence of tokens corresponding to the next item. However, uniformly applying our method to all tokens may hinder its effectiveness. Specifically, as more prefix tokens are established, the prediction for an item becomes increasingly certain, leading to reduced uncertainty in predicting subsequent tokens. Consequently, the difference between normal and counterfactual predictions—\textit{i.e.,} the causal effect of the behavior sequence—naturally diminishes on these later tokens. In other words, the prediction of these tokens is increasingly less influenced by the behavior sequence. In this case, uniformly forcing these diminishing causal effects (even approaching zero) to fit data labels becomes inappropriate. To tackle the challenge, we introduce a mechanism that adaptively assigns greater weights to earlier tokens when emphasizing the causal effects on predictions, forming our final fine-tuning method. Since the method centers on using counterfactuals to estimate and emphasize causal effects, we name it Counterfactual Fine-Tuning (CFT).

The main contributions of this work are summarized as follows:
\begin{itemize}[leftmargin=*]
    \item We highlight the issue of insufficient utilization of behavior sequences in existing LLM-based recommendation methods and provide an analysis from a causal perspective.
    
    \item We propose a novel counterfactual fine-tuning method to enhance behavior modeling in LLM-based recommendations by explicitly emphasizing the effects of behavior sequences on predictions during tuning. 
    
    \item We conduct extensive experiments on several real-world datasets, demonstrating the effectiveness of our method in enhancing behavior sequence modeling.
\end{itemize}

\section{Preliminary}
In this section, we present the problem formulation for the studied recommendation task and the background of the LLM-based recommendation methods this work focuses on.



\subsection{Next-item Recommendation}
Next-item prediction is the core task in recommender systems. In this work, we concentrate on this task within a sequential context. Let \(\mathcal{D}\) represent the collected historical data. We denote a sample in  $\mathcal{D}$ by \((u, h, y) \in \mathcal{D}\), where \(u\) represents a user, \(h\) denotes the user's historical interacted items (\ie, behavior sequence) up to a given time point, and \(y\) signifies the ground-truth next item interacted by the user. 
Notably, all the items could be represented by their textual information, primarily the title.  
Each user may have multiple samples in $\mathcal{D}$ by considering different interacted items as the next item $y$. 
Our objective is to train a model based on \(\mathcal{D}\), which can recommend appropriate items from the total item pool for the next-item interaction, given a user's historical behavior sequence.



\begin{table}
\centering
  \caption{
        Example of the instruction template for the Book dataset. The historical behavior sequence (or the ground-truth next item) would be placed in the \textit{<His\_Behavior\_Seq>} (or \textit{<Next\_Item>}) field.
    }
 \label{tab:instruct}
    \begin{tabularx}{0.45\textwidth}{lX}    
    \toprule
        \multicolumn{2}{c}{\textbf{Instruction Input}} \\
        \cdashline{1-2}[1pt/2.5pt]\noalign{\vskip 0.5ex} 
        \textbf{Task Instruction:}  & Given ten books that the user watched recently, please recommend a new book that the user likes to the user.\\ 
        \cdashline{1-2}[1pt/2.5pt]\noalign{\vskip 0.5ex} 
        
        \textbf{Task Input:} & The user has watched the following books before: \textit{<His\_Behavior\_Seq>} \\  
        \midrule
        \multicolumn{2}{c}{\textbf{Instruction Output}}  \\
        \cdashline{1-2}[1pt/2.5pt]\noalign{\vskip 0.5ex}
        \textbf{Output:}  & \textit{<Next\_Item>}  \\ 
    \bottomrule
    \end{tabularx}
\end{table}

\subsection{LLM-based Recommender} \label{sec:bigrec}
Many types of LLM-based recommendation methods have been developed. Among these, fine-tuning LLMs for recommendation in a generative manner is particularly well-suited for the next-item prediction task we defined, and it aligns more closely with the generative nature of LLMs. 
Next, we present the details of the tuning and inference processes for this type of approach, using a representative method BIGRec~\cite{bigrec} as an example.

\vspace{+5pt}
\noindent\textbf{Tuning}. To leverage LLMs for recommendation in a generative manner, this approach typically involves directly fine-tuning the LLMs to generate the next item. The first step is to convert each training example \((u,h,y) \in \mathcal{D}\) 
into the instruction format shown in Table~\ref{tab:instruct}, which consists of two parts: instruction input and instruction output. As indicated in the table, the user data (primarily \(h\)) and task instruction form the instruction input, denoted as \(x_{h}\), while the ground-truth next item \(y\) is directly treated as the instruction 
output. 
The LLM is then fine-tuned using this instruction data by optimizing the conditional language modeling objective. Formally, the optimization loss (denoted by $L_{n}$) can be formulated as follows:
\begin{equation}\label{eq:normalFT}
L_{n} =  \sum_{(u,h,y)\in \mathcal{D}} \sum_{t=1}^{|y|} \ell \big( f_{\theta}(x_{h},y_{<t}); \, y_t \big),
\end{equation}
where \(\ell\) denotes the Cross-Entropy loss, \(f_\theta (\cdot)\) denotes the LLM parameterized with $\theta$; $|y|$ denotes the total number of tokens for $y$, and \(y_t\) refers to the \(t\)-th token in \(y\). Notably, when predicting the \(t\)-th token, all preceding tokens in \(y\), denoted as \(y_{<t}\), are also used as input to the LLM, along with the instruction \(x_h\), to generate the prediction for $t$-th token, represented by \(f_\theta(x_h, y_{<t})\).




\vspace{+5pt}
\noindent\textbf{Inference}. After fine-tuning, the LLM is expected to have the ability to generate items as recommendations during the inference stage. However, since LLMs can generate creative content, potentially leading to generating nonexistent items. To solve the problem, BIGRec further considers performing a matching mechanism, finding the real items that are mostly similar to the generated ones as the final recommendation. The similarity is measured by the $L2$ distance between the generated item representations and the actual item representations encoded by the LLMs.


Notably, most methods in this sub-field share similar fine-tuning processes but differ in how they generate items at inference. For example, \(D^3\)~\cite{d3} additionally addresses the issue of the amplifying bias toward certain items during generation, and some other works consider directly rejecting non-actual items during the generation process. Since our method focuses on the tuning process, it is broadly applicable across these variations.

\section{Methodology}


In this section, we first conduct a causal analysis of the LLM prediction process to establish a foundation for our method design. We then introduce our CFT method, which explicitly emphasizes the effects of behavior sequences on predictions during training to enhance behavior modeling. 

\subsection{Causal Analysis}
We abstract the process of LLM prediction generation in the causal graph in Figure~\ref{fig:causal_graph}, in which nodes represent the involved variables and edges describe the causal relations between the nodes. We explain the causal graph as follows: 
\begin{itemize}[leftmargin=*]
    \item Node $Y_t$ denotes the prediction for the $t$-th token of next item.
    
    \item Node $H$ represents the historical behavior sequence in the input of the LLM.
    
    \item Node \(I\) represents all other input information, such as the task instruction and previously generated tokens (\(y_{<t}\) in Equation~\eqref{eq:normalFT}).
    
    \item Node $E$ represents the pre-training knowledge within LLMs.
    
    \item Path $\{H, I\}\rightarrow E \rightarrow Y_t$ represents that $H$ and $I$ can indirectly affect the prediction $Y_t$ through triggering the the pertaining knowledge within LLMs.

    \item Path $\{H, I\}\rightarrow Y_t$ represents that $H,I$ may also directly affect $Y_t$.
    
\end{itemize}

These paths illustrate how the model utilizes the inputs to produce results through different mechanisms.
Notably, the strength of these paths is dynamically learned through data fitting. 
However, they may encounter different learning challenges — the paths associated with the behavior sequence may present greater difficulties due to the inherent complexity of behavior patterns. Consequently, the model may not fully utilize these paths for prediction, misaligning their true roles in data generation, \textit{i.e.,} insufficiently leveraging the behavior sequence. To enhance the behavior sequence modeling, we need to enhance the effects of paths related to behavior, \textit{i.e.,} the effects of the behavior sequence, on model prediction. 

\vspace{+3pt}
\noindent \textbf{Causal effects of behavior sequence}. Based on the causal graph and causal inference theory~\cite{pearl2009causality}, the causal effect of a sample's behavior sequence \(h\) on the prediction \(Y_t\), conditioned on a given \(I\), can be expressed as follows:
\begin{equation}\label{eq:causaleffect}
\begin{split}
    P(Y_t|H=do(h),I) - P(Y_t|H=do(0),I) \\
    = P(Y_{t}|H=h,I) - P(Y_{t}|H=0,I),
\end{split}
\end{equation}
where $H=do(h)$ represents intervening $H$ as $h$, and $H=do(0)$ represents intervening $H$ as "None". $P(Y_{t}|H=h,I)$ denotes the normal predictions,
while \(P(Y_t|H=0, I)\) denotes the counterfactual result obtained by assuming the user has no historical behavior sequence.



\begin{figure}[t]
    \centering
    \includegraphics[width=0.49\textwidth]{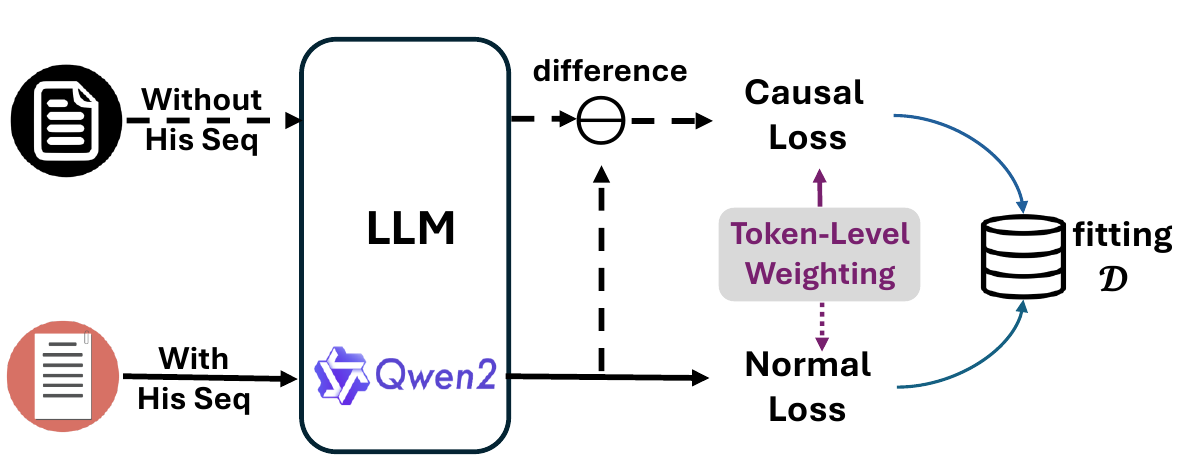}
    \caption{
    An overview of the proposed CFT framework, which includes two key components: a new task (the causal loss component) introduced in a multi-task manner and a token-level weighting mechanism. 
    }
    \label{fig:CFT}
    \Description{..}
\end{figure}

\subsection{Counterfactual Fine-Tuning}
Based on the causal analysis, we propose the Counterfactual Fine-Tuning (CFT) method, to enhance behavior sequence modeling in LLMs. CFT generally follows the paradigm of tuning LLMs to predict the next item but explicitly emphasizes the influence of behavior sequences on predictions during the training process, by introducing a new task. 
As shown in Figure~\ref{fig:CFT}, our approach consists of two main components:  
\begin{itemize}[leftmargin=*]
    \item \textbf{Multi-task Tuning}:  
    The core of our method lies in introducing a new task that directly uses the effect defined in Equation~\eqref{eq:causaleffect} to fit the training data. This new task emphasizes learning the effects of behavior sequences during data fitting, improving the model's utilization of behavior sequence information. We introduce this new task in a multi-task manner, retaining the original tuning task in Equation~\eqref{eq:normalFT} to preserve other valuable information. 
    
     
    \item \textbf{Token-level Weighting}:
     We apply a token-level weighting mechanism\footnote{The weighting mechanism is also optional for the original task. 
     } to adjust the strength of the new task loss, aligning the fact that behavior sequences have varying levels of influence on tokens at different positions. 
     

     
\end{itemize}


\subsubsection{Multi-task Tuning}
When fine-tuning LLMs, we additionally introduce a new task of directly using the effect of behavior sequence on model predictions to fit data, combining it with the task of using the norm predictions to fit data. Let $L_c$ denote the loss for our new task (termed \textit{causal loss}), and $L_n$ denote the loss for the original task (termed \textit{normal loss}). The multi-task tuning is performed by optimizing the following combined loss function:
\begin{equation}\label{eq:multitask}
L = L_n + \lambda L_c,    
\end{equation}
where $L$ denotes the combined loss, and $\lambda \geq0$ is a hyper-parameter to control the weight of the causal loss.

\vspace{+5pt}
\noindent \textbf{{Causal Loss $L_c$}}: 
The new task leverages the causal effect of behavior sequences to fit the data. To achieve this, we need to identify these effects, as outlined in Equation~\eqref{eq:causaleffect}. Since this equation is defined from a probabilistic perspective, we must convert it into an empirical form for practical application. For a sample \((u,h,y) \in \mathcal{D}\), the empirical representation of the effects for the \(t\)-th token prediction is given by:
\[
f_\theta(x_{h}, y_{<t}) - f_\theta(x_{0}, y_{<t}),
\]
where:
\begin{itemize}[leftmargin=*]
    \item[1)] \(f_{\theta}(x_{h}, y_{<t})\) represents the normal prediction for $y_t$, obtained by using the the instruction input \(x_h\) built on the user's behavior sequence. It corresponds to \(P(Y_t|H=h,I)\) in Equation~\eqref{eq:causaleffect}.
    \item[2)] \(f_{\theta}(x_{0}, y_{<t})\) represents the counterfactual prediction for $y_t$, corresponding to \(P(Y_t|H=0,I)\) in Equation~\eqref{eq:causaleffect}. It is obtained by assuming the user has no historical interactions, which means the <His\_Behavior\_Seq> field in the instruction template  (Table~\ref{tab:instruct}) is set to "None", forming the corresponding instruction input  \(x_{0}\) without the behavior sequence $h$.
\end{itemize}

After obtaining the effects, the causal loss \(L_c\) is formulated as:
\begin{equation}\label{eq:causalloss}
L_c =  \frac{1}{\Omega} \sum_{(u,h,y) \in \mathcal{D}} \sum_{t=1}^{|y|}  \omega_t \ell \big( f_{\theta}(x_h, y_{<t}) - f_{\theta}(x_0, y_{<t}); \, y_t \big),
\end{equation}
where \(\ell\) still denotes the Cross-Entropy loss, \(\omega_t\) represents the token-level weight that will be explained later, and \(\Omega\) is the sum of \(\omega_t\) across all predicted tokens. As shown in the equation, the task emphasizes attributing the occurrences of \(y_{t}\) to the behavior sequence's effects, thereby explicitly enhancing the utilization of the behavior sequence. 

\vspace{+5pt}
\noindent \textbf{{Normal Loss $L_n$}}: 
The other task still uses the normal prediction $f(x_{h}, y_{<t})$ to fit the data, as done by existing work described in Section~\ref{sec:bigrec}. So the loss $L_n$ can be computed following Equation~\eqref{eq:normalFT}.
A little differently, since the later tokens are easier to learn due to their lower uncertainty, we can also use a similar weighting mechanism to Equation~\eqref{eq:causalloss} to assign lower weights to these tokens during implementation.

\subsubsection{Token-level Weighting}
For a sample \((u,h,y)\in \mathcal{D}\), the behavior sequence should have varying levels of influence when predicting different position tokens in \(y\). As more prefix tokens are generated, the later item tokens become increasingly definitive by nature\footnote{For example, even for an untuned LLM, the prediction accuracy for the final tokens can reach an average of 0.74 (compared to just 0.05 for the first token) when considering only the prefix tokens.}, and even for some tokens, they may almost be entirely definitive given the prefix tokens. That means, the later tokens are less influenced by the behavior sequence and are primarily determined by the item's prefix tokens. In such cases, we should also make sure that, for the later tokens, the behavior sequence shows fewer effects during data fitting. Therefore, we design a token-level weighting mechanism 
that dynamically assigns decreasing weights from the first to the last item token on the corresponding loss.

Specifically, for each \(y\), we use a linear decay mechanism to set weights for item tokens based on their position. The first token of \(y\) is assigned the highest weight, set to 1, and the last token is assigned the lowest weight, set to \(\beta \,(\in [0,1]) \). The weight for fitting the \(t\)-th token in \(y\) is formulated as follows:
\begin{equation}\label{eq:weighting}
    \omega_t = 1 - \frac{(1-\beta) \cdot (t-1)}{|y|-1},
\end{equation}
where \(|y|\) represents the total number of tokens in \(y\), and \(\beta\) is a hyper-parameter controlling the lowest weight among the item's tokens. The weight decreases by \(\frac{1-\beta}{|y|-1}\) with each successive position. This weighting mechanism effectively assigned lower weights to the later tokens. The weights can be directly used by Equation~\eqref{eq:causalloss}, which naturally has a normalization mechanism. 

\begin{algorithm}[t]
	\caption{Counterfactual Fine-Tuning}
	\LinesNumbered
	\label{alg:CFT}
	\KwIn{Training data $\mathcal{D}$, hyper-parameters $\lambda$ and $\beta$}
    
    \While{Stop condition is not reached}{
    
    Compute the normal prediction $f_\theta(x_h,y_{<t})$ using the instruction with the behavior sequence\;

    Compute the counterfactual prediction $f_\theta(x_{0},y_{<t})$ by setting the historical sequences to "None"\;

    Use $f_\theta(x_h,y_{<t})$ as the prediction to compute the normal loss $L_n$ in Equation~\eqref{eq:normalFT}\;

    Use $f_\theta(x_h,y_{<t})-f_\theta(x_0,y_{<t})$ as the  causal effects to compute  the causal loss $L_c$ in Equation~\eqref{eq:causalloss}\;
    
    Update LLM parameters $\theta$ by optimizing the combined loss Equation~\eqref{eq:multitask}, \textit{i.e.,} $L=L_n + \lambda L_c$; 
		}
\end{algorithm}

\vspace{+5pt}
Algorithm~\ref{alg:CFT} outlines the pseudo-code for the tuning process in our CFT. In each iteration, we first compute the normal predictions \(f_{\theta}(x_h,y_t)\) and the counterfactual predictions \(f_{\theta}(x_0,y_t)\) (lines 2-3). Next, we calculate the normal loss using the normal predictions, as defined in Equation~\eqref{eq:normalFT} (line 4), and use the difference between the two predictions to determine the causal effect, which is then employed to compute the causal loss in Equation~\eqref{eq:causalloss} (line 5). Finally, we combine both losses to update the model parameters (line 6).

\vspace{+5pt}
\subsubsection{Inference.} Our proposed CFT just works at the fine-tuning stage for the LLM. Therefore, at inference stage, our method still uses the normal prediction to generate the recommendations, keeping the same as that described in Section~\ref{sec:bigrec}.






\section{Experiment}
In this section, we conduct a series of experiments to answer the
following research questions:

\noindent \textbf{RQ1}: How does CFT perform on real-world datasets compared to traditional and LLM-based sequential recommendation methods?

\noindent \textbf{RQ2}: What is the impact of the individual components of CFT on its effectiveness?



\noindent \bym{\textbf{RQ3}: How does CFT influences the recommendation distribution?}

\noindent \bym{\textbf{RQ4}: How does the backbone LLM choice and dataset selection influence the effectiveness of CFT?}

\subsection{Experimental Settings}
\subsubsection{Datasets}

We conduct experiments on three datasets from the Amazon Product Review benchmark~\cite{amazon}: CDs, Games, and Books. These datasets represent different domains and contain user interactions (reviews) on products from the Amazon platform, spanning from May 1996 to October 2018.

We fully follow the setting in $D^3$ paper~\cite{d3} to pre-process these datasets. Specifically, due to the high computational cost of training LLMs, we limit the data to interactions from a single year (October 2017 to October 2018). We then apply a 5-core filtering to ensure that each user/item has a minimum of 5 samples. Subsequently, we split the data into training, validation, and test sets based on the timestamps of the interactions, with an 8:1:1 ratio. 
This chronological partitioning ensures that the testing interactions occur after all training and validation interactions, thereby preventing information leakage~\cite{receval}. More preprocessing details could refer to the original paper of $D^3$~\cite{d3}. The summarized statistics of the processed datasets are presented in Table~\ref{exp:dataset}.



\begin{table}[t]
\caption{Statistical details of the evaluation datasets.}
\label{exp:dataset}
\begin{tabular}{cccc}
\hline
Dataset & \#User & \#Item & \#Interaction \\ \hline
CDs     & 21,347       &  14,239      &   185,855            \\
Games   & 34,089       &  11,037      &   252,015            \\
Books   &  67,708      &  41,722       &    853,747           \\ \hline
\end{tabular}
\end{table}


\begin{table*}[t]
\caption{Performance comparison of all methods on different-domain datasets with metrics NDCG@$K$ and HR@$K$. The best results are highlighted in bold, and an asterisk (*) denotes the incorporation of LLM embeddings for embedding initialization.}
\label{exp:main}
\resizebox{0.98\textwidth}{!}{
\begin{tabular}{c|l|ccccc|cc|cc}
\hline
Datasets                & Metrics & Caser  & GRU4Rec & SASRec & GRU4Rec* & SASRec* & BIGRec & $D^3$  & BIGRec+CFT      & $D^3$+CFT       \\ \hline
\multirow{4}{*}{CDs}   & NDCG@5  & 0.0161 & 0.0248  & 0.0477 & 0.0435   & 0.0418  & 0.0780 & 0.0906 & 0.0855          & \textbf{0.0972} \\
                       & HR@5    & 0.0224 & 0.0342  & 0.0647 & 0.0566   & 0.0561  & 0.1131 & 0.1243 & 0.1309          & \textbf{0.1417} \\
                       & NDCG@10 & 0.0193 & 0.0288  & 0.0535 & 0.0482   & 0.0478  & 0.0945 & 0.1103 & 0.1056          & \textbf{0.1203} \\
                       & HR@10   & 0.0485 & 0.0467  & 0.0824 & 0.0715   & 0.0745  & 0.1631 & 0.1826 & 0.1925          & \textbf{0.2116} \\ \hline
\multirow{4}{*}{Games} & NDCG@5  & 0.0122 & 0.0169  & 0.0237 & 0.0282   & 0.0263  & 0.0377 & 0.0432 & 0.0398          & \textbf{0.0453} \\
                       & HR@5    & 0.0187 & 0.0261  & 0.0338 & 0.0407   & 0.0383  & 0.0565 & 0.0631 & 0.0608          & \textbf{0.0674} \\
                       & NDCG@10 & 0.0164 & 0.0221  & 0.0290 & 0.0352   & 0.0328  & 0.0434 & 0.0498 & 0.0451          & \textbf{0.0516} \\
                       & HR@10   & 0.0321 & 0.0423  & 0.0502 & 0.0624   & 0.0586  & 0.0742 & 0.0836 & 0.0775          & \textbf{0.0868} \\ \hline
\multirow{4}{*}{Books} & NDCG@5  & 0.0042 & 0.0060  & 0.0097 & 0.0076   & 0.0103  & 0.0260 & 0.0261 & 0.0292          & \textbf{0.0298} \\
                       & HR@5    & 0.0069 & 0.0094  & 0.0146 & 0.0119   & 0.0128  & 0.0394 & 0.0375 & \textbf{0.0449} & 0.0435          \\
                       & NDCG@10 & 0.0060 & 0.0078  & 0.0123 & 0.0099   & 0.0151  & 0.0333 & 0.0329 & 0.0359          & \textbf{0.0363} \\
                       & HR@10   & 0.0123 & 0.0149  & 0.0226 & 0.0189   & 0.0229  & 0.0612 & 0.0585 & \textbf{0.0655}          & 0.0635          \\ \hline
\end{tabular}
}
\end{table*}


\subsubsection{Evaluation Settings}
To evaluate recommendation performance, we employ two widely recognized metrics: Hit Ratio (HR@K) and Normalized Discounted Cumulative Gain (NDCG@K), with \(K \in \{5, 10\}\). 
HR@$K$ measures whether the ground-truth item is included in the top-K recommendations, while NDCG@K assesses the ranking quality by considering the relative order of the ground-truth item within the top-K list. 
Higher values for both metrics indicate better performance.
In our evaluation, these metrics are computed using the all-ranking protocol~\cite{bigrec}, where all items that a user has not interacted with are treated as potential candidates. Additionally, during testing, the interactions immediately preceding the test interaction, including the one at testing sets, are included in the user's historical behavior sequence to input to the model, similar to prior work~\cite{d3}.


\subsubsection{Compared Methods}

To demonstrate the superiority of our method, we compare it against the following traditional sequential methods (Caser, GRU4Rec, SASRec) and LLM-based methods (BIGRec, $D^3$): 
\begin{itemize}[leftmargin=*]
    \item \textbf{Caser}~\cite{caser}. 
    This is a famous sequential recommendation approach that employs Convolutional Neural Networks (CNNs) to encode sequential patterns for modeling user preferences.
    \item \textbf{GRU4Rec}~\cite{gru4rec}. 
    This is another famous method that employs Gated Recurrent Units (GRU) to encode sequential patterns for modeling user preferences.

    \item \textbf{SASRec}~\cite{sasrec}. 
    This is a highly representative sequential recommendation method that employs the self-attention network for user preference modeling.

    \item \textbf{GRU4Rec*}~\cite{gru4rec}. This is a variant of GRU4Rec that initializes the item embeddings in GRU4Rec using those encoded by LLMs.
     
    \item \textbf{SASRec*}~\cite{sasrec}. This is a variant of SASRec that initializes embeddings in SASRec using those encoded by LLMs.
    
    \item \textbf{BIGRec}~\cite{bigrec}. This is a representative LLM-based recommendation method that fine-tunes LLMs to generate the next items based on input behavior sequences, as introduced in Section~\ref{sec:bigrec}.
    
    
    \item $\bm{D^3}$~\cite{d3}. 
    This is a state-of-the-art LLM-based recommendation method. It follows a similar fine-tuning process to BIGRec but differs during inference. Specifically, it mitigates the amplification bias toward certain items by removing length normalization in LLM beam search decoding. 
    Besides, it also includes an ensemble design with traditional models, but we omit this design in our implementation for a fair comparison.
\end{itemize}

For our method, we implement two variations by applying CFT for tuning while using the inference processes from BIGRec and \(D^3\). We refer to these implementations as BIGRec+CFT and \(D^3\)+CFT. 


\subsubsection{Implementing Details} \label{sec:implementation}


For all LLM-based methods compared, we use Qwen2-0.5B~\cite{qwen2} as the backbone LLM.  When tuning models, we use the AdamW~\cite{adamw} optimizer with a batch size of 64, a learning rate of \(1 \times 10^{-4}\), and a dropout rate of 0.05. Model selection is based on validation loss, using an early stopping strategy with a patience of one epoch. Other settings generally follow those in the $D^3$ paper.
For our CFT method's \(\lambda\), which controls the weight of the causal loss, is tuned in the range \{0.01, 0.02, 0.025, 0.05, 0.1, 0.2, 0.3\}. For our method's \(\beta\), which controls token-level weights, we introduce another hyper-parameter \(\beta'\) to facilitate implementation, where \(\beta = 1 - 1/\beta'\), and we tune \(\beta'\) within \{1.1, 1.2, 1.4, 1.6, 2, 3, 10, 25\}\footnote{Approximately equivalent to tuning \(\beta\) within \{0.09, 0.16, 0.29, 0.38, 0.5, 0.66, 0.9, 0.96\}}. 
Due to the high cost of tuning LLMs, we avoid grid search. Instead, we first identify the general scale of a hyper-parameter and then adjust it within a narrower range.
For all traditional methods, we strictly follow the settings in the $D^3$ paper~\cite{d3}.


For BIGRec's inference process, we adjust the original method, which generates a single item and matches it with actual items to form the top-K recommendation list. Instead, we generate five items. For each of these generated items, we find the most closely matched actual item and combine them to create the Top-5 recommendation list. We then identify the second-best matched items for each generated item and append them to the Top-5 list, resulting in a Top-10 recommendation list.
This approach helps avoid recommending overly similar items, significantly improving performance\footnote{Specifically, we saw an increase in NDCG@5 from 0.041 to 0.078 on the CDs dataset for BIGRec}. All LLM-based methods follow this implementation.

\subsection{Performance Comparison (RQ1)}
We begin by assessing the overall recommendation performance of the compared methods. The summarized results are presented in Table~\ref{exp:main}, where the results of traditional methods are sourced from the $D^3$ paper~\cite{d3}, from which we draw the following observations:

\begin{itemize}[leftmargin=*]
    
    \item One of our CFT implementations ($D^3$+CFT)  consistently outperforms the baselines across all evaluation metrics on all datasets. 
    This verifies the superiority of our CFT.

    \item 
    For LLM-based methods, the performance of BIGRec+CFT surpasses that of BIGRec, achieving an average relative improvement of 9.8\% across all metrics and datasets, while $D^3$+CFT surpasses $D^3$ with an average relative improvement of 9.5\%. This observation indicates the validity of our causal analysis, suggesting that existing methods may inadequately leverage behavior sequences, leading to sub-optimal performance. Furthermore, it highlights the effectiveness of CFT in enhancing behavior sequence modeling in LLMs by emphasizing the influence of behavior sequences on predictions.

    \item Traditional recommendation methods exhibit poor performance. Although incorporating LLM embeddings for initialization offers some improvements in most cases, a significant gap still exists compared to LLM-based recommendation methods. 
    This demonstrates the advantages of utilizing LLMs as recommendation models on these datasets. 
    

    
    


    \item In most cases, $D^3$ outperforms BIGRec, with only a slight decline in performance for some metrics on Books. This generally aligns with the observations in the $D^3$ paper (the version without ensemble). However, in our results, the performance improvements of BIGRec (and $D^3$) over traditional methods are significantly larger than those reported in the $D^3$ paper. This discrepancy can be attributed to our modification of the item-matching step during inference (see Section~\ref{sec:implementation}),
    which is more suitable for our setting.
    For instance, if using the original matching method in BIGRec, it fails to surpass SASRec in our setting (\textit{e.g.}, NDCG@5 on CDs: SASRec 0.0477 vs. BIGRec 0.0404).

\end{itemize}

\begin{table*}[t]
\caption{Ablation results for our proposed CFT on BIGRec, where `w/o CL', `w/o TW', and `w/o Both' indicate removing our new task, the token-level weighting mechanism, and both components, respectively. The metric NDCG@K is abbreviated as NG@K.}
\resizebox{1.0\textwidth}{!}{
\begin{tabular}{c|cccc|cccc|cccc}
\hline
Dataset & \multicolumn{4}{c|}{CDs} & \multicolumn{4}{c|}{Games} & \multicolumn{4}{c}{Books} \\ \hline
\multicolumn{1}{l|}{Metrics} & \multicolumn{1}{l}{NG@5} & \multicolumn{1}{l}{HR@5} & \multicolumn{1}{l}{NG@10} & \multicolumn{1}{l|}{HR@10} & \multicolumn{1}{l}{NG@5} & \multicolumn{1}{l}{HR@5} & \multicolumn{1}{l}{NG@10} & \multicolumn{1}{l|}{HR@10} & \multicolumn{1}{l}{NG@5} & \multicolumn{1}{l}{HR@5} & \multicolumn{1}{l}{NG@10} & \multicolumn{1}{l}{HR@10} \\ \hline
CFT & \textbf{0.0855} & \textbf{0.1309} & \textbf{0.1056} & \textbf{0.1925} & \textbf{0.0398} & \textbf{0.0608} & \textbf{0.0451} & \textbf{0.0775} & \textbf{0.0292} & \textbf{0.0449} & \textbf{0.0359} & \textbf{0.0655} \\
w/o CL & 0.0793 & 0.1184 & 0.0961 & 0.1702 & 0.0334 & 0.0492 & 0.0378 & 0.0628 & 0.0274 & 0.0411 & 0.0338 & 0.0607 \\
w/o TW & 0.0827 & 0.1264 & 0.1028 & 0.1884 & 0.0384 & 0.0578 & 0.0437 & 0.0742 & 0.0260 & 0.0393 & 0.0330 & 0.0612 \\
w/o Both & 0.0780 & 0.1131 & 0.0945 & 0.1632 & 0.0377 & 0.0565 & 0.0434 & 0.0742 & 0.0260 & 0.0394 & 0.0333 & 0.0612 \\ \hline
\end{tabular}
}
\label{exp:abl}
\end{table*}

\subsection{Ablation Study (RQ2)}
To enhance behavior sequence modeling in LLM-based Recommendations, CFT includes two key designs: a new task (using the causal effects of behavior sequences to fit data) and a token-level weighting mechanism.
To validate the rationale behind these design decisions, we conduct a comprehensive evaluation by systematically disabling each component of BIGRec+CFT to create several variants. Specifically, the following variants are introduced:
\begin{itemize}[leftmargin=*]
    \item \textbf{w/o CL}. 
    This variant disables the new task by setting the weight of the causal loss in Equation~\eqref{eq:multitask} to zero. Notably, the token-level weighting mechanism is still applied to the normal loss. 
    
    
    \item \textbf{w/o TW}. 
    This variant disables the token-level weighting mechanism independently, which is equivalent to setting the hyper-parameter $\beta$ in Equation~\eqref{eq:weighting} to one.
    
    \item \textbf{w/o Both}. 
    This variant removes both components mentioned above, which is equivalent to the vanilla BIGRec method.
    
\end{itemize}

Table~\ref{exp:abl} illustrates the ablation results of the BIGRec+CFT method, from which
we draw the following observations:
\begin{itemize}[leftmargin=*]
    \item Removing the new task (w/o CL) leads to a significant performance decline, confirming the impact of the introduced causal loss and underscoring the importance of emphasizing the influence of behavior sequences on model prediction
    for enhancing the utilization of behavior sequences. 
    %

    \item Disabling the token-level weighting mechanism (w/o TW) also results in a performance decline, confirming that it plays a crucial role in fully unlocking the potential of our method, by aligning the fact that behavior sequences have varying levels of influence on different tokens.
    

    \item Comparing the impact of disabling the new task (w/o CL) versus disabling the token-level weighting mechanism (w/o TW), disabling the new task results in a much more significant performance decline, indicating that the new task plays a more fundamental role in our method. 

    
    \item
    Comparing the variant w/o CL and the variant w/o Both, the w/o CL variant, which applies token-level weighting to the normal loss, still brings some improvements in most cases. This verifies that for the normal task, different tokens may also have different learning difficulties during tuning.


\end{itemize}

These results demonstrate that leveraging the effects of behavior sequences to fit data is central to our method; however, fully unlocking its potential also depends on integrating other designs.

\subsection{In-depth Analysis (RQ3 \& RQ4)}

In this subsection, we first analyze how CFT influences the recommendation distribution by comparing it to BIGRec, answering RQ3. Then, we investigate the impact of the backbone LLM choice and dataset selection on CFT's effectiveness, addressing RQ4.

\begin{figure}[t]
\centering
\subfigure[\textbf{ With History on Books}]{\includegraphics[width=0.23\textwidth]{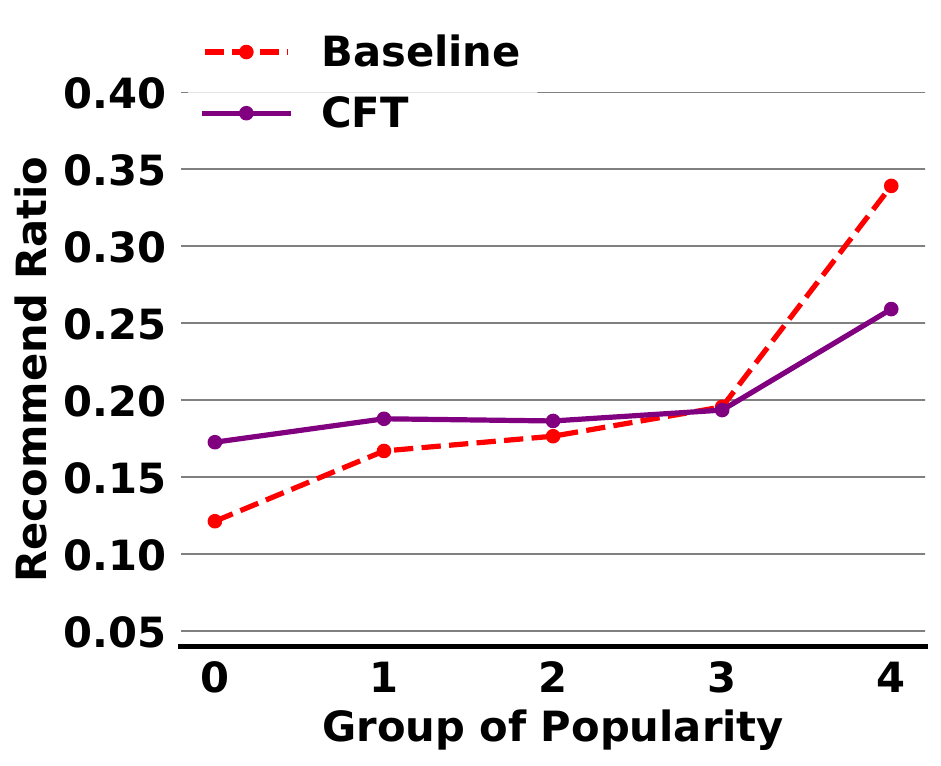}}
\subfigure[\textbf{ Without History on Books}]{\includegraphics[width=0.23\textwidth]{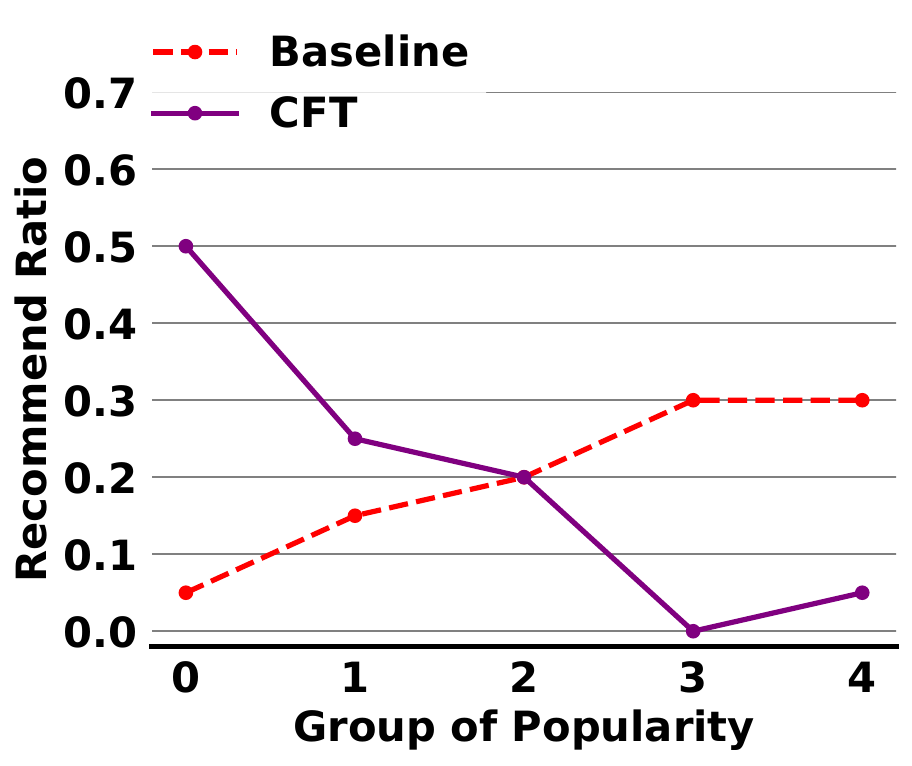}}
\subfigure[ \textbf{ With History on Games}]{ \includegraphics[width=0.23\textwidth]{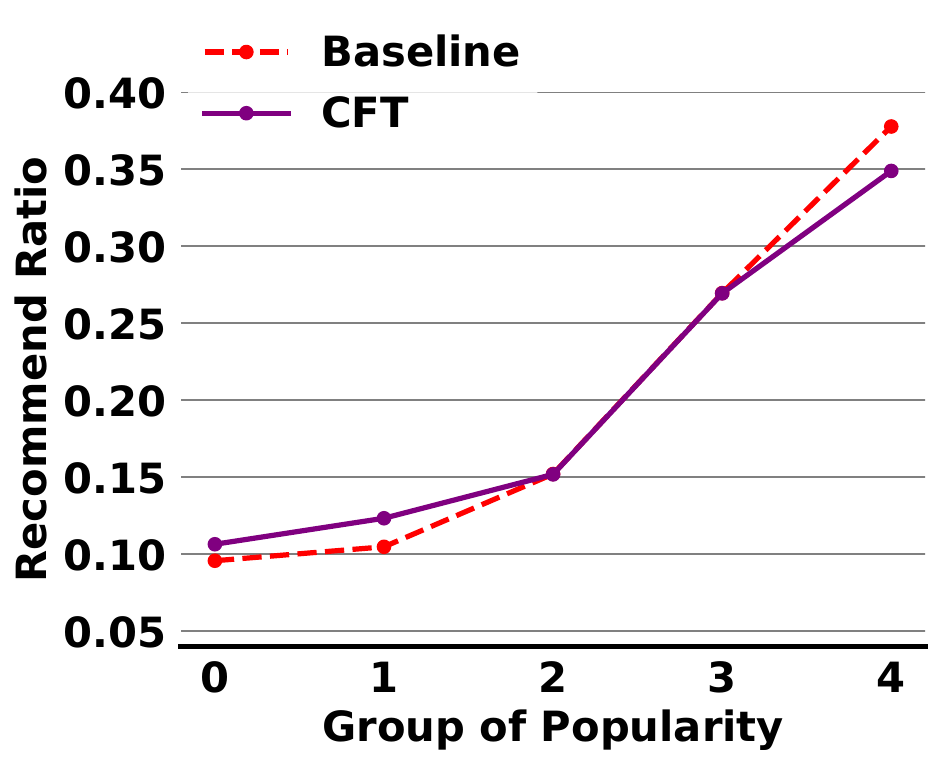}}
\subfigure[ \textbf{ Without History on Games}]{ \includegraphics[width=0.23\textwidth]{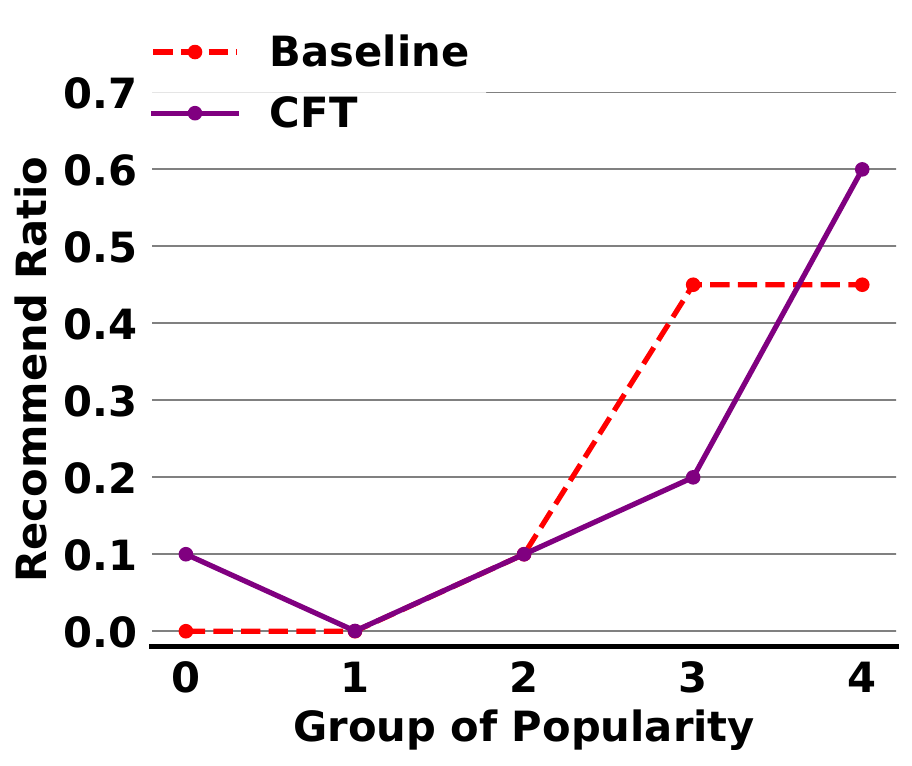}}
\caption{Top-20 recommendation distribution comparison between BIGRec (Baseline) and BIGRec + CFT (CFT).}
\label{fig:cmp-cft}

\Description{..}
\end{figure}

\subsubsection{Recommendation List Analysis}

We first conduct a study to analyze the impact of CFT on LLM recommendations. To do this, we compare the distribution of recommended items between BIGRec and our CFT implemented on BIGRec (BIGRec+CFT). Specifically, we categorize items into groups based on their popularity and calculate the proportion of recommendations each group receives in the final recommendation list, that generated by BIGRec and BIGRec+CFT in the case with and without inputting behavior sequences. We draw the comparison results in Figure~\ref{fig:cmp-cft}, where the item group with higher popularity has a higher index. From the figure, we draw the following observations:
\begin{itemize}[leftmargin=*]
    \item Focusing on comparing recommendations generated by CFT and BIGRec when inputting behavior sequence, we find that CFT can produce more balanced recommendations among the different item groups --- reducing the recommendation to the popular items and increasing the recommendation towards the unpopular items. This aligns with an intuition that --- if the behavior sequence (personalization information) has not been fully utilized, the model may tend to recommend common popular items after tuning.  The result somewhat shows that our method can leverage the behavior sequence more.

    \item 
    Focusing on cases without historical behavior sequences, 
    we observe that CFT introduces notable changes compared to BIGRec. In particular, for the Book dataset, when behavior input is absent, CFT shifts towards recommending a large number of unpopular items that users are less likely to consume, significantly misaligning with the results when full behavior input is provided. This demonstrates that CFT reduces the model's reliance on non-behavioral knowledge when making recommendations.   

 
\end{itemize}

\begin{table}[]
\caption{Performance comparison on the LLaMA-3.2 backbone across CDs and Books datasets.}
\begin{tabular}{c|l|ccc}
\hline
{\color[HTML]{333333} Dataset}                 & {\color[HTML]{333333} Metrics}          & {\color[HTML]{333333} SASRec} & {\color[HTML]{333333} BIGRec}          & {\color[HTML]{333333} BIGRec+CFT}             \\ \hline
{\color[HTML]{333333} }                        & {\color[HTML]{333333} NDCG@5}  & {\color[HTML]{333333} 0.0477} & {\color[HTML]{333333} \textbf{0.0673}} & {\color[HTML]{333333} 0.0672}          \\
{\color[HTML]{333333} }                        & {\color[HTML]{333333} HR@5}    & {\color[HTML]{333333} 0.0647} & {\color[HTML]{333333} 0.0907}          & {\color[HTML]{333333} \textbf{0.0947}} \\
{\color[HTML]{333333} }                        & {\color[HTML]{333333} NDCG@10} & {\color[HTML]{333333} 0.0535} & {\color[HTML]{333333} 0.0801}          & {\color[HTML]{333333} \textbf{0.0825}} \\
\multirow{-4}{*}{{\color[HTML]{333333} CDs}}   & {\color[HTML]{333333} HR@10}   & {\color[HTML]{333333} 0.0824} & {\color[HTML]{333333} 0.1292}          & {\color[HTML]{333333} \textbf{0.1413}} \\ \hline
{\color[HTML]{333333} }                        & {\color[HTML]{333333} NDCG@5}           & {\color[HTML]{333333} 0.0097} & {\color[HTML]{333333} 0.0229}          & {\color[HTML]{333333} \textbf{0.0239}} \\
{\color[HTML]{333333} }                        & {\color[HTML]{333333} HR@5}             & {\color[HTML]{333333} 0.0146} & {\color[HTML]{333333} 0.0325}          & {\color[HTML]{333333} \textbf{0.0369}} \\
{\color[HTML]{333333} }                        & {\color[HTML]{333333} NDCG@10}          & {\color[HTML]{333333} 0.0123} & {\color[HTML]{333333} 0.0295}          & {\color[HTML]{333333} \textbf{0.0308}} \\
\multirow{-4}{*}{{\color[HTML]{333333} Books}} & {\color[HTML]{333333} HR@10}            & {\color[HTML]{333333} 0.0226} & {\color[HTML]{333333} 0.0529}          & {\color[HTML]{333333} \textbf{0.0583}} \\ \hline
\end{tabular}
\label{exp:llama}
\end{table}

\subsubsection{Method Effectiveness on Other LLM Backbones} 
Next, we assess the effectiveness of our method using a different LLM backbone, Llama3.2-1B~\cite{llama3}. For comparison, we include SASRec, BIGRec, and our CFT implemented on BIGRec (BIGRec+CFT). Regarding datasets, we select the Books and CDs datasets, as they represent the cases where CFT showed the highest and lowest relative improvements over BIGRec in the main results. To minimize overhead, we keep the hyper-parameters (e.g., learning rate, dropout) consistent with those in the main experiment. Table~\ref{exp:llama} summarizes the results. As shown, both BIGRec and BIGRec+CFT outperform the traditional method SASRec. Moreover, our CFT continues to improve BIGRec across all cases, except for NDCG@5 on CDs. Specifically, CFT achieves an average relative improvement of 4.2\% on CDs and 8.1\% on Books, demonstrating that our method can be effectively applied to other LLM backbones.

\subsubsection{Method Effectiveness on Datasets beyond Amazon} 
\begin{figure}[t]
    \centering
    \subfigure
    {\includegraphics[width=0.23\textwidth]{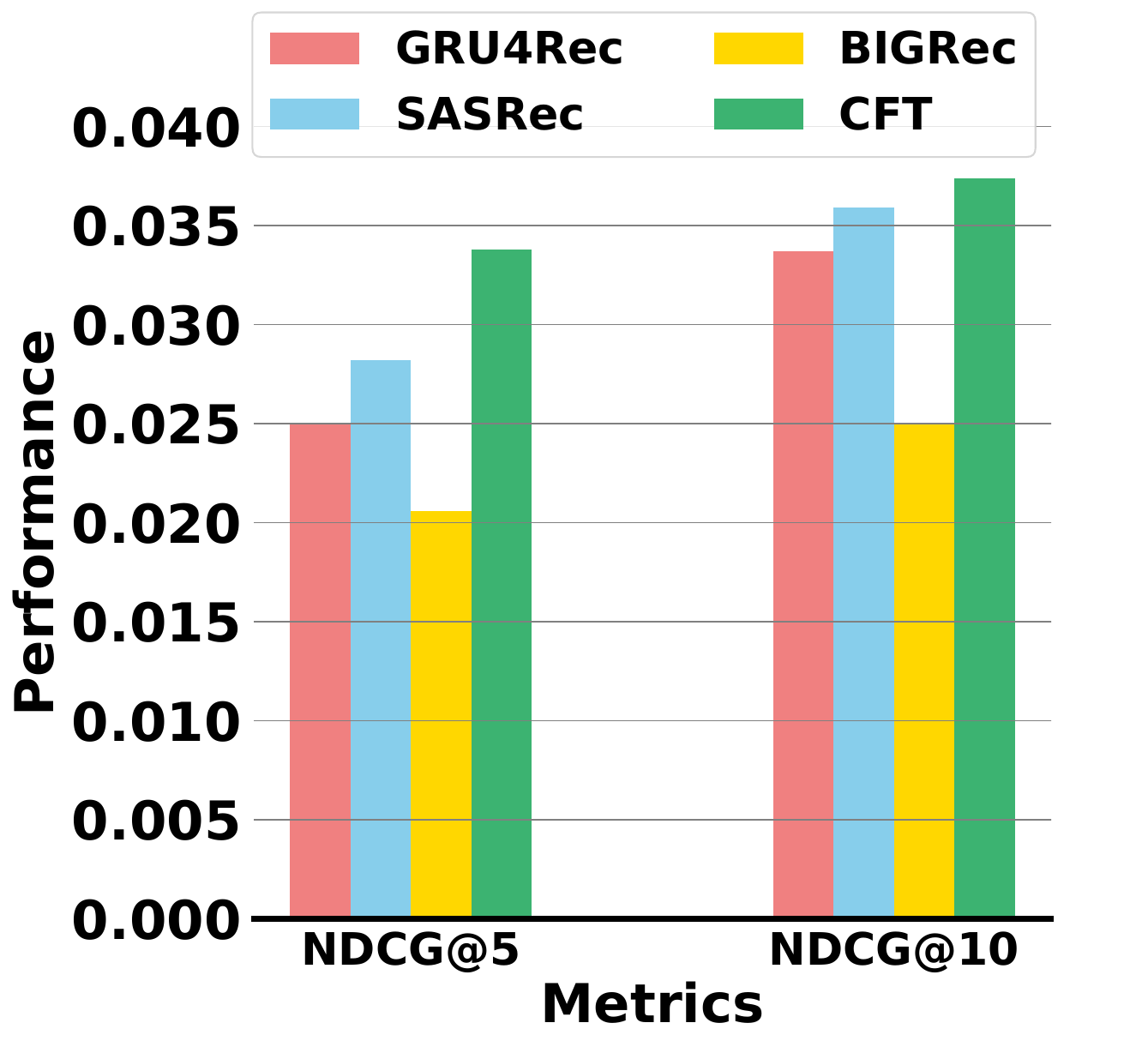}}
    \subfigure
    {\includegraphics[width=0.23\textwidth]{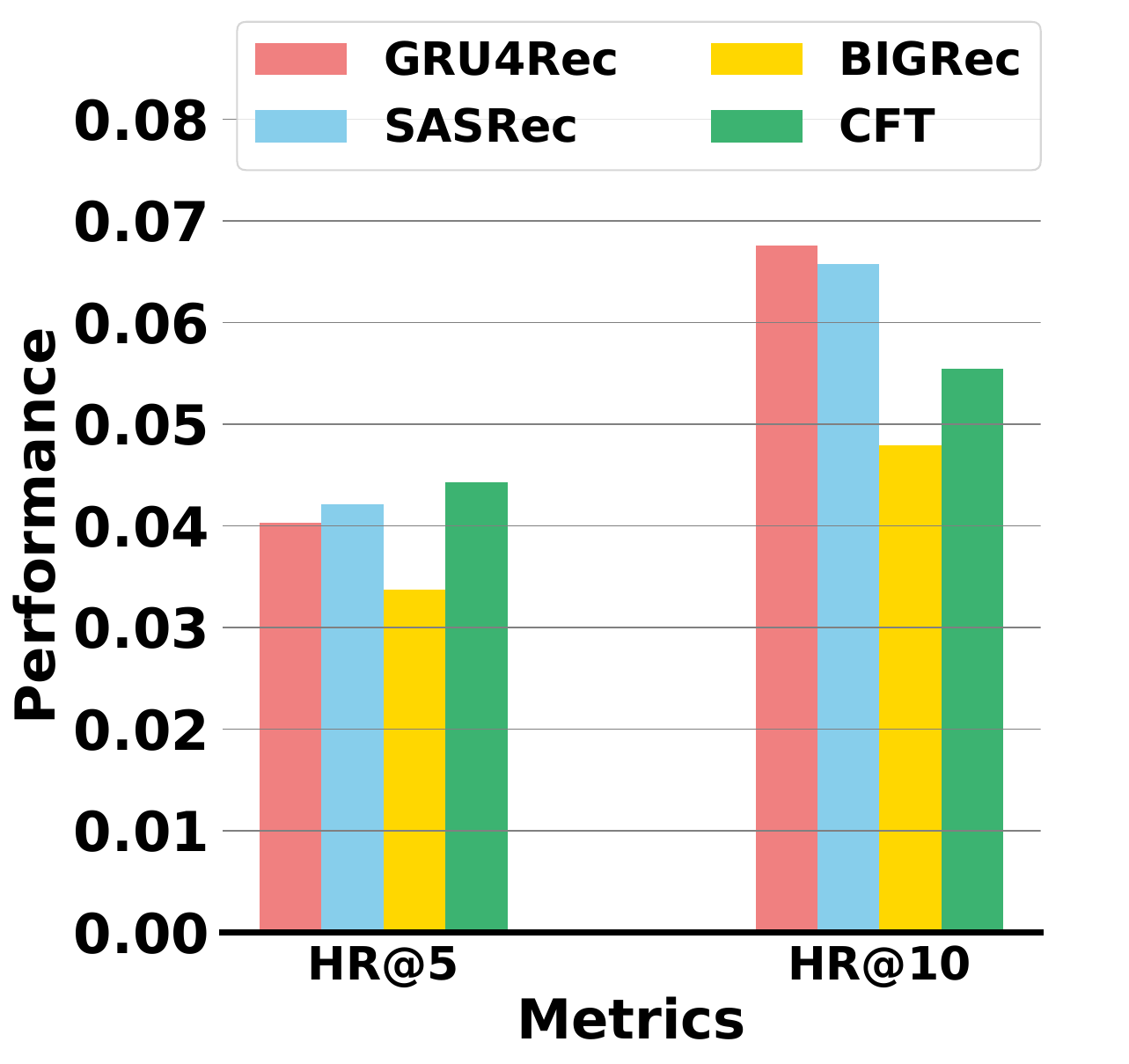}}
    \caption{Performance comparison on Steam dataset. 
    }
    \label{fig:steam}
    \Description{..}
\end{figure}

In the main experiment, we assessed the method's effectiveness using datasets from Amazon. Here, we conduct further studies using the Steam~\cite{steam} dataset, comparing GRU4Rec, SASRec, and BIGRec against our CFT (implemented based on BIGRec), as shown in Figure~\ref{fig:steam}. 
Our method consistently achieves the best results in 3 out of 4 cases. On average across all metrics, our CFT demonstrates a relative improvement of 40.2\% over BIGRec and a 3.4\% improvement over the best traditional baselines. Notably, BIGRec does not outperform the traditional baselines on this Steam dataset; however, when applying our CFT, it surpasses these baselines (except for the HR@10 metric). This further validates the effectiveness of our CFT.


\section{Related Work}
In this section, we discuss related work on LLM-based recommendation and causal recommendation.

\subsection{LLM-based Recommendation}

Given the significant and widespread success of large language models (LLMs), the recommendation community has expressed great enthusiasm for adapting LLMs to recommendation tasks. Current explorations can be divided into three main categories: 1) optimizing prompts or leveraging in-context learning to inspire the capabilities of LLMs for recommendation better~\cite{llm4isr,zero-shot_next_item,chatrec}; 2) employing an agent paradigm to utilize the planning and reasoning abilities of LLMs for recommendations~\cite{agentcf,gen_agent,recmind}; and 3) tuning LLMs based on recommendation data to align them with the recommendation task, enhancing their recommendation abilities via model updates~\cite{tallrec,bigrec,binllm}. Among these approaches, tuning methods have garnered the most attention and are more relevant to this paper, thereby we mainly discuss this type of method.

Regarding the research of tuning, early research predominantly focused on a discriminative approach~\cite{tallrec,nijianmo}, where candidates are provided to LLMs to assess user preferences. This method has certain drawbacks, particularly due to the high costs associated with all-ranking~\cite{bigrec}.
To better leverage the generative capabilities of LLMs, some studies have emerged that directly tune large models to generate items, including works like BIGRec~\cite{bigrec} and GPT4Rec~\cite{gpt4rec}. Following these two lines of research, new exploration directions have arisen that address the problems that have more recommendation characteristics. For instance, some studies investigate how to better incorporate collaborative information into large model recommendations~\cite{cllm4rec, binllm, lcrec}, and some studies focus on how to better represent items within LLMs~\cite{idgenrec, transrec,letter}. Additionally, there are efforts aimed at developing decoding methods suitable for LLMs~\cite{d3} or addressing challenges related to long-sequence modeling~\cite{trsr,rella}, or accelerating LLM-based recommendations~\cite{xi2024decoding,lin2024efficient}.
However, to our knowledge, we are the first to utilize causality to enhance the behavior sequence modeling for LLMs.



\subsection{Causal Recommendation}
Causality has a long history of application in recommendations, primarily focusing on addressing bias issues~\cite{gao2024causal,causal_infer}. Initially, inverse propensity scores were widely employed for debiasing, where the core idea is to adjust the training distribution to be unbiased by reweighting training samples with propensity scores~\cite{depce,li2023propensity,saito2020unbiased}. Subsequently, causal interventions based on do-calculus have been utilized to tackle various bias problems brought by the existence of confounders, such as popularity bias~\cite{pda,causer}, duration bias~\cite{d2q}, confounding features~\cite{dcr}, and amplification bias~\cite{wenjie_deconfounding}. Additionally, some studies leverage counterfactual inference to address bias issues;
for instance, CR~\cite{clickbait} and CVRDD~\cite{cvrdd} utilize counterfactuals to tackle clickbait and duration bias, respectively.
All these works are centered around traditional recommender systems. Our research significantly differs from theirs. 
First, we specifically focus on LLM-based recommendations. Second, we address the insufficient utilization of behavior sequences, a challenge encountered during the fine-tuning of LLMs for recommendations, rather than the bias issues explored in prior research. From a technical perspective, our approach significantly diverges, as it is tailored to LLMs, incorporating token-level weighting into the design.

\section{Conclusion}
In this work, we demonstrated that the existing LLM-based recommendation methods may suffer from the issue of insufficient utilization of behavior sequences. We provided a causal analysis of this problem and proposed a Counterfactual Fine-Tuning (CFT) method to enhance behavior sequence modeling. The core of our CFT approach involves introducing a new task that leverages the effects of behavior sequences to directly align with data labels. With a token-level weighting mechanism, the task could help explicitly emphasize the role of behavior sequences in model predictions. Extensive results validated the effectiveness of our method.

In current experiments, we focused exclusively on the LLM-based paradigm of tuning models to generate the next items based on textual information. In the future, we will explore our method within other frameworks, such as tuning LLMs to generate matching scores. 
We also plan to explore the issue in scenarios where additional personalization information—beyond behavior sequences, such as encoded collaborative embeddings~\cite{collm}—is utilized.



\bibliographystyle{ACM-Reference-Format}
\balance
\bibliography{7_reference}


\begin{thebibliography}{53}


\ifx \showCODEN    \undefined \def \showCODEN     #1{\unskip}     \fi
\ifx \showDOI      \undefined \def \showDOI       #1{#1}\fi
\ifx \showISBNx    \undefined \def \showISBNx     #1{\unskip}     \fi
\ifx \showISBNxiii \undefined \def \showISBNxiii  #1{\unskip}     \fi
\ifx \showISSN     \undefined \def \showISSN      #1{\unskip}     \fi
\ifx \showLCCN     \undefined \def \showLCCN      #1{\unskip}     \fi
\ifx \shownote     \undefined \def \shownote      #1{#1}          \fi
\ifx \showarticletitle \undefined \def \showarticletitle #1{#1}   \fi
\ifx \showURL      \undefined \def \showURL       {\relax}        \fi
\providecommand\bibfield[2]{#2}
\providecommand\bibinfo[2]{#2}
\providecommand\natexlab[1]{#1}
\providecommand\showeprint[2][]{arXiv:#2}

\bibitem[Bao et~al\mbox{.}(2023a)]%
        {bigrec}
\bibfield{author}{\bibinfo{person}{Keqin Bao}, \bibinfo{person}{Jizhi Zhang}, \bibinfo{person}{Wenjie Wang}, \bibinfo{person}{Yang Zhang}, \bibinfo{person}{Zhengyi Yang}, \bibinfo{person}{Yancheng Luo}, \bibinfo{person}{Chong Chen}, \bibinfo{person}{Fuli Feng}, {and} \bibinfo{person}{Qi Tian}.} \bibinfo{year}{2023}\natexlab{a}.
\newblock \showarticletitle{A bi-step grounding paradigm for large language models in recommendation systems}.
\newblock \bibinfo{journal}{\emph{arXiv preprint arXiv:2308.08434}} (\bibinfo{year}{2023}).
\newblock


\bibitem[Bao et~al\mbox{.}(2024)]%
        {d3}
\bibfield{author}{\bibinfo{person}{Keqin Bao}, \bibinfo{person}{Jizhi Zhang}, \bibinfo{person}{Yang Zhang}, \bibinfo{person}{Xinyue Huo}, \bibinfo{person}{Chong Chen}, {and} \bibinfo{person}{Fuli Feng}.} \bibinfo{year}{2024}\natexlab{}.
\newblock \showarticletitle{Decoding matters: Addressing amplification bias and homogeneity issue for llm-based recommendation}.
\newblock \bibinfo{journal}{\emph{EMNLP}} (\bibinfo{year}{2024}).
\newblock


\bibitem[Bao et~al\mbox{.}(2023b)]%
        {tallrec}
\bibfield{author}{\bibinfo{person}{Keqin Bao}, \bibinfo{person}{Jizhi Zhang}, \bibinfo{person}{Yang Zhang}, \bibinfo{person}{Wenjie Wang}, \bibinfo{person}{Fuli Feng}, {and} \bibinfo{person}{Xiangnan He}.} \bibinfo{year}{2023}\natexlab{b}.
\newblock \showarticletitle{Tallrec: An effective and efficient tuning framework to align large language model with recommendation}. In \bibinfo{booktitle}{\emph{Proceedings of the 17th ACM Conference on Recommender Systems}}. \bibinfo{pages}{1007--1014}.
\newblock


\bibitem[Dubey et~al\mbox{.}(2024)]%
        {llama3}
\bibfield{author}{\bibinfo{person}{Abhimanyu Dubey}, \bibinfo{person}{Abhinav Jauhri}, \bibinfo{person}{Abhinav Pandey}, \bibinfo{person}{Abhishek Kadian}, \bibinfo{person}{Ahmad Al-Dahle}, \bibinfo{person}{Aiesha Letman}, \bibinfo{person}{Akhil Mathur}, \bibinfo{person}{Alan Schelten}, \bibinfo{person}{Amy Yang}, \bibinfo{person}{Angela Fan}, {et~al\mbox{.}}} \bibinfo{year}{2024}\natexlab{}.
\newblock \showarticletitle{The llama 3 herd of models}.
\newblock \bibinfo{journal}{\emph{arXiv preprint arXiv:2407.21783}} (\bibinfo{year}{2024}).
\newblock


\bibitem[Elkahky et~al\mbox{.}(2015)]%
        {user_modeling}
\bibfield{author}{\bibinfo{person}{Ali~Mamdouh Elkahky}, \bibinfo{person}{Yang Song}, {and} \bibinfo{person}{Xiaodong He}.} \bibinfo{year}{2015}\natexlab{}.
\newblock \showarticletitle{A multi-view deep learning approach for cross domain user modeling in recommendation systems}. In \bibinfo{booktitle}{\emph{Proceedings of the 24th international conference on world wide web}}. \bibinfo{pages}{278--288}.
\newblock


\bibitem[Gao et~al\mbox{.}(2024)]%
        {gao2024causal}
\bibfield{author}{\bibinfo{person}{Chen Gao}, \bibinfo{person}{Yu Zheng}, \bibinfo{person}{Wenjie Wang}, \bibinfo{person}{Fuli Feng}, \bibinfo{person}{Xiangnan He}, {and} \bibinfo{person}{Yong Li}.} \bibinfo{year}{2024}\natexlab{}.
\newblock \showarticletitle{Causal inference in recommender systems: A survey and future directions}.
\newblock \bibinfo{journal}{\emph{ACM Transactions on Information Systems}} \bibinfo{volume}{42}, \bibinfo{number}{4} (\bibinfo{year}{2024}), \bibinfo{pages}{1--32}.
\newblock


\bibitem[Gao et~al\mbox{.}(2023)]%
        {chatrec}
\bibfield{author}{\bibinfo{person}{Yunfan Gao}, \bibinfo{person}{Tao Sheng}, \bibinfo{person}{Youlin Xiang}, \bibinfo{person}{Yun Xiong}, \bibinfo{person}{Haofen Wang}, {and} \bibinfo{person}{Jiawei Zhang}.} \bibinfo{year}{2023}\natexlab{}.
\newblock \showarticletitle{Chat-rec: Towards interactive and explainable llms-augmented recommender system}.
\newblock \bibinfo{journal}{\emph{arXiv preprint arXiv:2303.14524}} (\bibinfo{year}{2023}).
\newblock


\bibitem[Gupta et~al\mbox{.}(2021)]%
        {causer}
\bibfield{author}{\bibinfo{person}{Priyanka Gupta}, \bibinfo{person}{Ankit Sharma}, \bibinfo{person}{Pankaj Malhotra}, \bibinfo{person}{Lovekesh Vig}, {and} \bibinfo{person}{Gautam Shroff}.} \bibinfo{year}{2021}\natexlab{}.
\newblock \showarticletitle{Causer: Causal session-based recommendations for handling popularity bias}. In \bibinfo{booktitle}{\emph{Proceedings of the 30th ACM international conference on information \& knowledge management}}. \bibinfo{pages}{3048--3052}.
\newblock


\bibitem[He et~al\mbox{.}(2023)]%
        {dcr}
\bibfield{author}{\bibinfo{person}{Xiangnan He}, \bibinfo{person}{Yang Zhang}, \bibinfo{person}{Fuli Feng}, \bibinfo{person}{Chonggang Song}, \bibinfo{person}{Lingling Yi}, \bibinfo{person}{Guohui Ling}, {and} \bibinfo{person}{Yongdong Zhang}.} \bibinfo{year}{2023}\natexlab{}.
\newblock \showarticletitle{Addressing Confounding Feature Issue for Causal Recommendation}.
\newblock \bibinfo{journal}{\emph{ACM Trans. Inf. Syst.}} \bibinfo{volume}{41}, \bibinfo{number}{3}, Article \bibinfo{articleno}{53} (\bibinfo{date}{Feb.} \bibinfo{year}{2023}), \bibinfo{numpages}{23}~pages.
\newblock
\showISSN{1046-8188}
\urldef\tempurl%
\url{https://doi.org/10.1145/3559757}
\showDOI{\tempurl}


\bibitem[Hidasi et~al\mbox{.}(2016)]%
        {gru4rec}
\bibfield{author}{\bibinfo{person}{Bal{\'{a}}zs Hidasi}, \bibinfo{person}{Alexandros Karatzoglou}, \bibinfo{person}{Linas Baltrunas}, {and} \bibinfo{person}{Domonkos Tikk}.} \bibinfo{year}{2016}\natexlab{}.
\newblock \showarticletitle{Session-based Recommendations with Recurrent Neural Networks}. In \bibinfo{booktitle}{\emph{4th International Conference on Learning Representations}}.
\newblock


\bibitem[Hou et~al\mbox{.}(2024)]%
        {zoranker}
\bibfield{author}{\bibinfo{person}{Yupeng Hou}, \bibinfo{person}{Junjie Zhang}, \bibinfo{person}{Zihan Lin}, \bibinfo{person}{Hongyu Lu}, \bibinfo{person}{Ruobing Xie}, \bibinfo{person}{Julian McAuley}, {and} \bibinfo{person}{Wayne~Xin Zhao}.} \bibinfo{year}{2024}\natexlab{}.
\newblock \showarticletitle{Large language models are zero-shot rankers for recommender systems}. In \bibinfo{booktitle}{\emph{European Conference on Information Retrieval}}. Springer, \bibinfo{pages}{364--381}.
\newblock


\bibitem[Ji et~al\mbox{.}(2023)]%
        {receval}
\bibfield{author}{\bibinfo{person}{Yitong Ji}, \bibinfo{person}{Aixin Sun}, \bibinfo{person}{Jie Zhang}, {and} \bibinfo{person}{Chenliang Li}.} \bibinfo{year}{2023}\natexlab{}.
\newblock \showarticletitle{A critical study on data leakage in recommender system offline evaluation}.
\newblock \bibinfo{journal}{\emph{ACM Transactions on Information Systems}} \bibinfo{volume}{41}, \bibinfo{number}{3} (\bibinfo{year}{2023}), \bibinfo{pages}{1--27}.
\newblock


\bibitem[Kang and McAuley(2018)]%
        {sasrec}
\bibfield{author}{\bibinfo{person}{Wang-Cheng Kang} {and} \bibinfo{person}{Julian McAuley}.} \bibinfo{year}{2018}\natexlab{}.
\newblock \showarticletitle{Self-attentive sequential recommendation}. In \bibinfo{booktitle}{\emph{2018 IEEE international conference on data mining (ICDM)}}. IEEE, \bibinfo{pages}{197--206}.
\newblock


\bibitem[Kang et~al\mbox{.}(2023)]%
        {nijianmo}
\bibfield{author}{\bibinfo{person}{Wang-Cheng Kang}, \bibinfo{person}{Jianmo Ni}, \bibinfo{person}{Nikhil Mehta}, \bibinfo{person}{Maheswaran Sathiamoorthy}, \bibinfo{person}{Lichan Hong}, \bibinfo{person}{Ed Chi}, {and} \bibinfo{person}{Derek~Zhiyuan Cheng}.} \bibinfo{year}{2023}\natexlab{}.
\newblock \showarticletitle{Do llms understand user preferences? evaluating llms on user rating prediction}.
\newblock \bibinfo{journal}{\emph{arXiv preprint arXiv:2305.06474}} (\bibinfo{year}{2023}).
\newblock


\bibitem[Li et~al\mbox{.}(2023a)]%
        {li2023propensity}
\bibfield{author}{\bibinfo{person}{Haoxuan Li}, \bibinfo{person}{Yanghao Xiao}, \bibinfo{person}{Chunyuan Zheng}, \bibinfo{person}{Peng Wu}, {and} \bibinfo{person}{Peng Cui}.} \bibinfo{year}{2023}\natexlab{a}.
\newblock \showarticletitle{Propensity matters: Measuring and enhancing balancing for recommendation}. In \bibinfo{booktitle}{\emph{International Conference on Machine Learning}}. PMLR, \bibinfo{pages}{20182--20194}.
\newblock


\bibitem[Li et~al\mbox{.}(2023c)]%
        {gpt4rec}
\bibfield{author}{\bibinfo{person}{Jinming Li}, \bibinfo{person}{Wentao Zhang}, \bibinfo{person}{Tian Wang}, \bibinfo{person}{Guanglei Xiong}, \bibinfo{person}{Alan Lu}, {and} \bibinfo{person}{Gerard Medioni}.} \bibinfo{year}{2023}\natexlab{c}.
\newblock \showarticletitle{GPT4Rec: A generative framework for personalized recommendation and user interests interpretation}.
\newblock \bibinfo{journal}{\emph{arXiv preprint arXiv:2304.03879}} (\bibinfo{year}{2023}).
\newblock


\bibitem[Li et~al\mbox{.}(2023b)]%
        {prompt_distill}
\bibfield{author}{\bibinfo{person}{Lei Li}, \bibinfo{person}{Yongfeng Zhang}, {and} \bibinfo{person}{Li Chen}.} \bibinfo{year}{2023}\natexlab{b}.
\newblock \showarticletitle{Prompt distillation for efficient llm-based recommendation}. In \bibinfo{booktitle}{\emph{Proceedings of the 32nd ACM International Conference on Information and Knowledge Management}}. \bibinfo{pages}{1348--1357}.
\newblock


\bibitem[Liang et~al\mbox{.}(2016)]%
        {causal_infer}
\bibfield{author}{\bibinfo{person}{Dawen Liang}, \bibinfo{person}{Laurent Charlin}, {and} \bibinfo{person}{David~M Blei}.} \bibinfo{year}{2016}\natexlab{}.
\newblock \showarticletitle{Causal inference for recommendation}. In \bibinfo{booktitle}{\emph{Causation: Foundation to Application, Workshop at UAI. AUAI}}, Vol.~\bibinfo{volume}{6}. \bibinfo{pages}{108}.
\newblock


\bibitem[Lin et~al\mbox{.}(2024a)]%
        {llmrec_survey}
\bibfield{author}{\bibinfo{person}{Jianghao Lin}, \bibinfo{person}{Xinyi Dai}, \bibinfo{person}{Yunjia Xi}, \bibinfo{person}{Weiwen Liu}, \bibinfo{person}{Bo Chen}, \bibinfo{person}{Hao Zhang}, \bibinfo{person}{Yong Liu}, \bibinfo{person}{Chuhan Wu}, \bibinfo{person}{Xiangyang Li}, \bibinfo{person}{Chenxu Zhu}, \bibinfo{person}{Huifeng Guo}, \bibinfo{person}{Yong Yu}, \bibinfo{person}{Ruiming Tang}, {and} \bibinfo{person}{Weinan Zhang}.} \bibinfo{year}{2024}\natexlab{a}.
\newblock \showarticletitle{How Can Recommender Systems Benefit from Large Language Models: A Survey}.
\newblock \bibinfo{journal}{\emph{ACM Trans. Inf. Syst.}} (\bibinfo{date}{July} \bibinfo{year}{2024}).
\newblock
\showISSN{1046-8188}
\urldef\tempurl%
\url{https://doi.org/10.1145/3678004}
\showDOI{\tempurl}
\newblock
\shownote{Just Accepted}.


\bibitem[Lin et~al\mbox{.}(2024b)]%
        {transrec}
\bibfield{author}{\bibinfo{person}{Xinyu Lin}, \bibinfo{person}{Wenjie Wang}, \bibinfo{person}{Yongqi Li}, \bibinfo{person}{Fuli Feng}, \bibinfo{person}{See-Kiong Ng}, {and} \bibinfo{person}{Tat-Seng Chua}.} \bibinfo{year}{2024}\natexlab{b}.
\newblock \showarticletitle{Bridging Items and Language: A Transition Paradigm for Large Language Model-Based Recommendation}. In \bibinfo{booktitle}{\emph{Proceedings of the 30th ACM SIGKDD Conference on Knowledge Discovery and Data Mining}}. \bibinfo{pages}{1816--1826}.
\newblock


\bibitem[Lin et~al\mbox{.}(2024c)]%
        {lin2024efficient}
\bibfield{author}{\bibinfo{person}{Xinyu Lin}, \bibinfo{person}{Chaoqun Yang}, \bibinfo{person}{Wenjie Wang}, \bibinfo{person}{Yongqi Li}, \bibinfo{person}{Cunxiao Du}, \bibinfo{person}{Fuli Feng}, \bibinfo{person}{See-Kiong Ng}, {and} \bibinfo{person}{Tat-Seng Chua}.} \bibinfo{year}{2024}\natexlab{c}.
\newblock \showarticletitle{Efficient Inference for Large Language Model-based Generative Recommendation}.
\newblock \bibinfo{journal}{\emph{arXiv preprint arXiv:2410.05165}} (\bibinfo{year}{2024}).
\newblock


\bibitem[Loshchilov and Hutter(2019)]%
        {adamw}
\bibfield{author}{\bibinfo{person}{Ilya Loshchilov} {and} \bibinfo{person}{Frank Hutter}.} \bibinfo{year}{2019}\natexlab{}.
\newblock \showarticletitle{Decoupled Weight Decay Regularization}. In \bibinfo{booktitle}{\emph{International Conference on Learning Representations}}.
\newblock
\urldef\tempurl%
\url{https://openreview.net/forum?id=Bkg6RiCqY7}
\showURL{%
\tempurl}


\bibitem[Ngo and Nguyen(2024)]%
        {recgpt}
\bibfield{author}{\bibinfo{person}{Hoang Ngo} {and} \bibinfo{person}{Dat~Quoc Nguyen}.} \bibinfo{year}{2024}\natexlab{}.
\newblock \showarticletitle{RecGPT: Generative Pre-training for Text-based Recommendation}. In \bibinfo{booktitle}{\emph{Proceedings of the 62nd Annual Meeting of the Association for Computational Linguistics, {ACL} 2024 - Student Research Workshop, Bangkok, Thailand, August 11-16, 2024}}. \bibinfo{publisher}{Association for Computational Linguistics}, \bibinfo{pages}{302--313}.
\newblock


\bibitem[Ni et~al\mbox{.}(2019)]%
        {amazon}
\bibfield{author}{\bibinfo{person}{Jianmo Ni}, \bibinfo{person}{Jiacheng Li}, {and} \bibinfo{person}{Julian McAuley}.} \bibinfo{year}{2019}\natexlab{}.
\newblock \showarticletitle{Justifying Recommendations using Distantly-Labeled Reviews and Fine-Grained Aspects}. In \bibinfo{booktitle}{\emph{Proceedings of the 2019 Conference on Empirical Methods in Natural Language Processing and the 9th International Joint Conference on Natural Language Processing (EMNLP-IJCNLP)}}. \bibinfo{pages}{188--197}.
\newblock


\bibitem[Pearl(2009)]%
        {pearl2009causality}
\bibfield{author}{\bibinfo{person}{J Pearl}.} \bibinfo{year}{2009}\natexlab{}.
\newblock \bibinfo{booktitle}{\emph{Causality}}.
\newblock \bibinfo{publisher}{Cambridge university press}.
\newblock


\bibitem[Pearl(2016)]%
        {pearl2016primer}
\bibfield{author}{\bibinfo{person}{Judea Pearl}.} \bibinfo{year}{2016}\natexlab{}.
\newblock \bibinfo{booktitle}{\emph{Causal Inference in Statistics: A Primer}}.
\newblock \bibinfo{publisher}{John Wiley \& Sons}.
\newblock


\bibitem[Rappaz et~al\mbox{.}(2021)]%
        {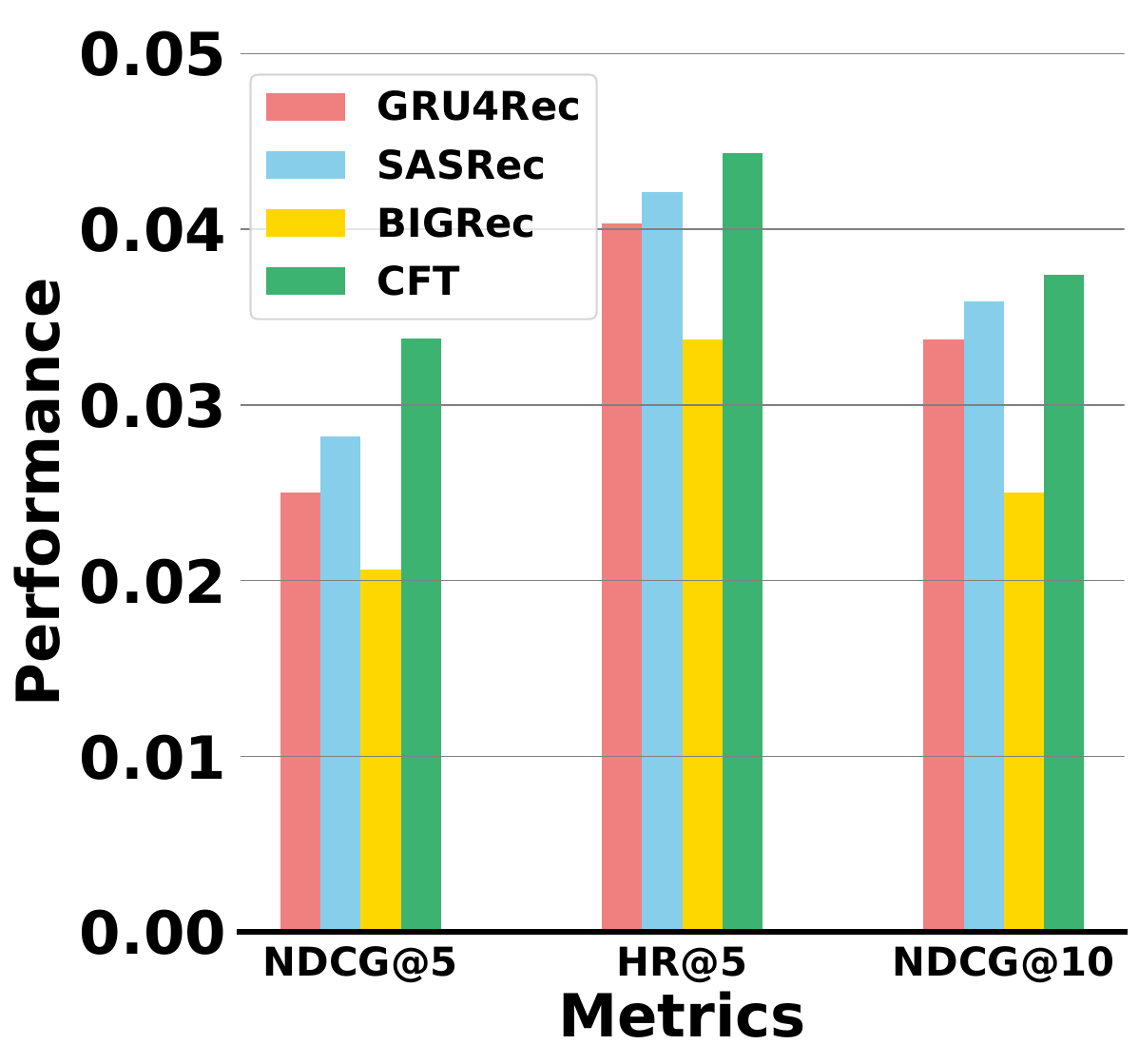}
\bibfield{author}{\bibinfo{person}{J\'{e}r\'{e}mie Rappaz}, \bibinfo{person}{Julian McAuley}, {and} \bibinfo{person}{Karl Aberer}.} \bibinfo{year}{2021}\natexlab{}.
\newblock \showarticletitle{Recommendation on Live-Streaming Platforms: Dynamic Availability and Repeat Consumption}. In \bibinfo{booktitle}{\emph{Proceedings of the 15th ACM Conference on Recommender Systems}} (Amsterdam, Netherlands) \emph{(\bibinfo{series}{RecSys '21})}. \bibinfo{publisher}{Association for Computing Machinery}, \bibinfo{address}{New York, NY, USA}, \bibinfo{pages}{390–399}.
\newblock
\showISBNx{9781450384582}
\urldef\tempurl%
\url{https://doi.org/10.1145/3460231.3474267}
\showDOI{\tempurl}


\bibitem[Saito et~al\mbox{.}(2020)]%
        {saito2020unbiased}
\bibfield{author}{\bibinfo{person}{Yuta Saito}, \bibinfo{person}{Suguru Yaginuma}, \bibinfo{person}{Yuta Nishino}, \bibinfo{person}{Hayato Sakata}, {and} \bibinfo{person}{Kazuhide Nakata}.} \bibinfo{year}{2020}\natexlab{}.
\newblock \showarticletitle{Unbiased recommender learning from missing-not-at-random implicit feedback}. In \bibinfo{booktitle}{\emph{Proceedings of the 13th International Conference on Web Search and Data Mining}}. \bibinfo{pages}{501--509}.
\newblock


\bibitem[Sun et~al\mbox{.}(2024)]%
        {llm4isr}
\bibfield{author}{\bibinfo{person}{Zhu Sun}, \bibinfo{person}{Hongyang Liu}, \bibinfo{person}{Xinghua Qu}, \bibinfo{person}{Kaidong Feng}, \bibinfo{person}{Yan Wang}, {and} \bibinfo{person}{Yew~Soon Ong}.} \bibinfo{year}{2024}\natexlab{}.
\newblock \showarticletitle{Large Language Models for Intent-Driven Session Recommendations}. In \bibinfo{booktitle}{\emph{Proceedings of the 47th International ACM SIGIR Conference on Research and Development in Information Retrieval}} (Washington DC, USA) \emph{(\bibinfo{series}{SIGIR '24})}. \bibinfo{publisher}{Association for Computing Machinery}, \bibinfo{address}{New York, NY, USA}, \bibinfo{pages}{324–334}.
\newblock
\showISBNx{9798400704314}
\urldef\tempurl%
\url{https://doi.org/10.1145/3626772.3657688}
\showDOI{\tempurl}


\bibitem[Tan et~al\mbox{.}(2024)]%
        {idgenrec}
\bibfield{author}{\bibinfo{person}{Juntao Tan}, \bibinfo{person}{Shuyuan Xu}, \bibinfo{person}{Wenyue Hua}, \bibinfo{person}{Yingqiang Ge}, \bibinfo{person}{Zelong Li}, {and} \bibinfo{person}{Yongfeng Zhang}.} \bibinfo{year}{2024}\natexlab{}.
\newblock \showarticletitle{IDGenRec: LLM-RecSys Alignment with Textual ID Learning}. In \bibinfo{booktitle}{\emph{Proceedings of the 47th International ACM SIGIR Conference on Research and Development in Information Retrieval}}. \bibinfo{pages}{355--364}.
\newblock


\bibitem[Tang and Wang(2018)]%
        {caser}
\bibfield{author}{\bibinfo{person}{Jiaxi Tang} {and} \bibinfo{person}{Ke Wang}.} \bibinfo{year}{2018}\natexlab{}.
\newblock \showarticletitle{Personalized Top-N Sequential Recommendation via Convolutional Sequence Embedding}. In \bibinfo{booktitle}{\emph{Proceedings of the Eleventh ACM International Conference on Web Search and Data Mining}} (Marina Del Rey, CA, USA) \emph{(\bibinfo{series}{WSDM '18})}. \bibinfo{publisher}{Association for Computing Machinery}, \bibinfo{address}{New York, NY, USA}, \bibinfo{numpages}{9}~pages.
\newblock
\showISBNx{9781450355810}
\urldef\tempurl%
\url{https://doi.org/10.1145/3159652.3159656}
\showDOI{\tempurl}


\bibitem[Tang et~al\mbox{.}(2023)]%
        {cvrdd}
\bibfield{author}{\bibinfo{person}{Shisong Tang}, \bibinfo{person}{Qing Li}, \bibinfo{person}{Dingmin Wang}, \bibinfo{person}{Ci Gao}, \bibinfo{person}{Wentao Xiao}, \bibinfo{person}{Dan Zhao}, \bibinfo{person}{Yong Jiang}, \bibinfo{person}{Qian Ma}, {and} \bibinfo{person}{Aoyang Zhang}.} \bibinfo{year}{2023}\natexlab{}.
\newblock \showarticletitle{Counterfactual Video Recommendation for Duration Debiasing}. In \bibinfo{booktitle}{\emph{Proceedings of the 29th ACM SIGKDD Conference on Knowledge Discovery and Data Mining}} (Long Beach, CA, USA) \emph{(\bibinfo{series}{KDD '23})}. \bibinfo{publisher}{Association for Computing Machinery}, \bibinfo{address}{New York, NY, USA}, \bibinfo{pages}{4894–4903}.
\newblock
\showISBNx{9798400701030}
\urldef\tempurl%
\url{https://doi.org/10.1145/3580305.3599797}
\showDOI{\tempurl}


\bibitem[Wang and Lim(2023)]%
        {zero-shot_next_item}
\bibfield{author}{\bibinfo{person}{Lei Wang} {and} \bibinfo{person}{Ee-Peng Lim}.} \bibinfo{year}{2023}\natexlab{}.
\newblock \showarticletitle{Zero-shot next-item recommendation using large pretrained language models}.
\newblock \bibinfo{journal}{\emph{arXiv preprint arXiv:2304.03153}} (\bibinfo{year}{2023}).
\newblock


\bibitem[Wang et~al\mbox{.}(2024a)]%
        {letter}
\bibfield{author}{\bibinfo{person}{Wenjie Wang}, \bibinfo{person}{Honghui Bao}, \bibinfo{person}{Xinyu Lin}, \bibinfo{person}{Jizhi Zhang}, \bibinfo{person}{Yongqi Li}, \bibinfo{person}{Fuli Feng}, \bibinfo{person}{See-Kiong Ng}, {and} \bibinfo{person}{Tat-Seng Chua}.} \bibinfo{year}{2024}\natexlab{a}.
\newblock \showarticletitle{Learnable Item Tokenization for Generative Recommendation}. In \bibinfo{booktitle}{\emph{Proceedings of the 33rd ACM International Conference on Information and Knowledge Management}}. \bibinfo{pages}{2400–2409}.
\newblock


\bibitem[Wang et~al\mbox{.}(2021a)]%
        {wenjie_deconfounding}
\bibfield{author}{\bibinfo{person}{Wenjie Wang}, \bibinfo{person}{Fuli Feng}, \bibinfo{person}{Xiangnan He}, \bibinfo{person}{Xiang Wang}, {and} \bibinfo{person}{Tat-Seng Chua}.} \bibinfo{year}{2021}\natexlab{a}.
\newblock \showarticletitle{Deconfounded recommendation for alleviating bias amplification}. In \bibinfo{booktitle}{\emph{Proceedings of the 27th ACM SIGKDD conference on knowledge discovery \& data mining}}. \bibinfo{pages}{1717--1725}.
\newblock


\bibitem[Wang et~al\mbox{.}(2021b)]%
        {clickbait}
\bibfield{author}{\bibinfo{person}{Wenjie Wang}, \bibinfo{person}{Fuli Feng}, \bibinfo{person}{Xiangnan He}, \bibinfo{person}{Hanwang Zhang}, {and} \bibinfo{person}{Tat-Seng Chua}.} \bibinfo{year}{2021}\natexlab{b}.
\newblock \showarticletitle{Clicks can be Cheating: Counterfactual Recommendation for Mitigating Clickbait Issue}. In \bibinfo{booktitle}{\emph{Proceedings of the 44th International ACM SIGIR Conference on Research and Development in Information Retrieval}} (Virtual Event, Canada) \emph{(\bibinfo{series}{SIGIR '21})}. \bibinfo{publisher}{Association for Computing Machinery}, \bibinfo{address}{New York, NY, USA}, \bibinfo{pages}{1288–1297}.
\newblock
\showISBNx{9781450380379}
\urldef\tempurl%
\url{https://doi.org/10.1145/3404835.3462962}
\showDOI{\tempurl}


\bibitem[Wang et~al\mbox{.}(2024b)]%
        {recmind}
\bibfield{author}{\bibinfo{person}{Yancheng Wang}, \bibinfo{person}{Ziyan Jiang}, \bibinfo{person}{Zheng Chen}, \bibinfo{person}{Fan Yang}, \bibinfo{person}{Yingxue Zhou}, \bibinfo{person}{Eunah Cho}, \bibinfo{person}{Xing Fan}, \bibinfo{person}{Yanbin Lu}, \bibinfo{person}{Xiaojiang Huang}, {and} \bibinfo{person}{Yingzhen Yang}.} \bibinfo{year}{2024}\natexlab{b}.
\newblock \showarticletitle{{R}ec{M}ind: Large Language Model Powered Agent For Recommendation}. In \bibinfo{booktitle}{\emph{Findings of the Association for Computational Linguistics: NAACL 2024}}. \bibinfo{publisher}{Association for Computational Linguistics}, \bibinfo{pages}{4351--4364}.
\newblock


\bibitem[Wei et~al\mbox{.}(2024)]%
        {llm4trdrec}
\bibfield{author}{\bibinfo{person}{Wei Wei}, \bibinfo{person}{Xubin Ren}, \bibinfo{person}{Jiabin Tang}, \bibinfo{person}{Qinyong Wang}, \bibinfo{person}{Lixin Su}, \bibinfo{person}{Suqi Cheng}, \bibinfo{person}{Junfeng Wang}, \bibinfo{person}{Dawei Yin}, {and} \bibinfo{person}{Chao Huang}.} \bibinfo{year}{2024}\natexlab{}.
\newblock \showarticletitle{Llmrec: Large language models with graph augmentation for recommendation}. In \bibinfo{booktitle}{\emph{Proceedings of the 17th ACM International Conference on Web Search and Data Mining}}. \bibinfo{pages}{806--815}.
\newblock


\bibitem[Xi et~al\mbox{.}(2024a)]%
        {rella}
\bibfield{author}{\bibinfo{person}{Yunjia Xi}, \bibinfo{person}{Weiwen Liu}, \bibinfo{person}{Jianghao Lin}, \bibinfo{person}{Xiaoling Cai}, \bibinfo{person}{Hong Zhu}, \bibinfo{person}{Jieming Zhu}, \bibinfo{person}{Bo Chen}, \bibinfo{person}{Ruiming Tang}, \bibinfo{person}{Weinan Zhang}, {and} \bibinfo{person}{Yong Yu}.} \bibinfo{year}{2024}\natexlab{a}.
\newblock \showarticletitle{Towards Open-World Recommendation with Knowledge Augmentation from Large Language Models}. In \bibinfo{booktitle}{\emph{Proceedings of the 18th ACM Conference on Recommender Systems}} \emph{(\bibinfo{series}{RecSys '24})}. \bibinfo{publisher}{Association for Computing Machinery}, \bibinfo{address}{New York, NY, USA}, \bibinfo{pages}{12–22}.
\newblock
\showISBNx{9798400705052}


\bibitem[Xi et~al\mbox{.}(2024b)]%
        {xi2024decoding}
\bibfield{author}{\bibinfo{person}{Yunjia Xi}, \bibinfo{person}{Hangyu Wang}, \bibinfo{person}{Bo Chen}, \bibinfo{person}{Jianghao Lin}, \bibinfo{person}{Menghui Zhu}, \bibinfo{person}{Weiwen Liu}, \bibinfo{person}{Ruiming Tang}, \bibinfo{person}{Weinan Zhang}, {and} \bibinfo{person}{Yong Yu}.} \bibinfo{year}{2024}\natexlab{b}.
\newblock \showarticletitle{A Decoding Acceleration Framework for Industrial Deployable LLM-based Recommender Systems}.
\newblock \bibinfo{journal}{\emph{arXiv preprint arXiv:2408.05676}} (\bibinfo{year}{2024}).
\newblock


\bibitem[Xu et~al\mbox{.}(2022)]%
        {depce}
\bibfield{author}{\bibinfo{person}{Chen Xu}, \bibinfo{person}{Jun Xu}, \bibinfo{person}{Xu Chen}, \bibinfo{person}{Zhenghua Dong}, {and} \bibinfo{person}{Ji-Rong Wen}.} \bibinfo{year}{2022}\natexlab{}.
\newblock \showarticletitle{Dually Enhanced Propensity Score Estimation in Sequential Recommendation}. In \bibinfo{booktitle}{\emph{Proceedings of the 31st ACM International Conference on Information \& Knowledge Management}} (Atlanta, GA, USA) \emph{(\bibinfo{series}{CIKM '22})}. \bibinfo{publisher}{Association for Computing Machinery}, \bibinfo{address}{New York, NY, USA}, \bibinfo{pages}{2260–2269}.
\newblock
\showISBNx{9781450392365}
\urldef\tempurl%
\url{https://doi.org/10.1145/3511808.3557299}
\showDOI{\tempurl}


\bibitem[Yang et~al\mbox{.}(2024)]%
        {qwen2}
\bibfield{author}{\bibinfo{person}{An Yang}, \bibinfo{person}{Baosong Yang}, \bibinfo{person}{Binyuan Hui}, \bibinfo{person}{Bo Zheng}, \bibinfo{person}{Bowen Yu}, \bibinfo{person}{Chang Zhou}, \bibinfo{person}{Chengpeng Li}, \bibinfo{person}{Chengyuan Li}, \bibinfo{person}{Dayiheng Liu}, \bibinfo{person}{Fei Huang}, \bibinfo{person}{Guanting Dong}, \bibinfo{person}{Haoran Wei}, \bibinfo{person}{Huan Lin}, \bibinfo{person}{Jialong Tang}, \bibinfo{person}{Jialin Wang}, \bibinfo{person}{Jian Yang}, \bibinfo{person}{Jianhong Tu}, \bibinfo{person}{Jianwei Zhang}, \bibinfo{person}{Jianxin Ma}, \bibinfo{person}{Jin Xu}, \bibinfo{person}{Jingren Zhou}, \bibinfo{person}{Jinze Bai}, \bibinfo{person}{Jinzheng He}, \bibinfo{person}{Junyang Lin}, \bibinfo{person}{Kai Dang}, \bibinfo{person}{Keming Lu}, \bibinfo{person}{Keqin Chen}, \bibinfo{person}{Kexin Yang}, \bibinfo{person}{Mei Li}, \bibinfo{person}{Mingfeng Xue}, \bibinfo{person}{Na Ni}, \bibinfo{person}{Pei Zhang}, \bibinfo{person}{Peng Wang}, \bibinfo{person}{Ru
  Peng}, \bibinfo{person}{Rui Men}, \bibinfo{person}{Ruize Gao}, \bibinfo{person}{Runji Lin}, \bibinfo{person}{Shijie Wang}, \bibinfo{person}{Shuai Bai}, \bibinfo{person}{Sinan Tan}, \bibinfo{person}{Tianhang Zhu}, \bibinfo{person}{Tianhao Li}, \bibinfo{person}{Tianyu Liu}, \bibinfo{person}{Wenbin Ge}, \bibinfo{person}{Xiaodong Deng}, \bibinfo{person}{Xiaohuan Zhou}, \bibinfo{person}{Xingzhang Ren}, \bibinfo{person}{Xinyu Zhang}, \bibinfo{person}{Xipin Wei}, \bibinfo{person}{Xuancheng Ren}, \bibinfo{person}{Yang Fan}, \bibinfo{person}{Yang Yao}, \bibinfo{person}{Yichang Zhang}, \bibinfo{person}{Yu Wan}, \bibinfo{person}{Yunfei Chu}, \bibinfo{person}{Yuqiong Liu}, \bibinfo{person}{Zeyu Cui}, \bibinfo{person}{Zhenru Zhang}, {and} \bibinfo{person}{Zhihao Fan}.} \bibinfo{year}{2024}\natexlab{}.
\newblock \showarticletitle{Qwen2 Technical Report}.
\newblock \bibinfo{journal}{\emph{arXiv preprint arXiv:2407.10671}} (\bibinfo{year}{2024}).
\newblock


\bibitem[Zhan et~al\mbox{.}(2022)]%
        {d2q}
\bibfield{author}{\bibinfo{person}{Ruohan Zhan}, \bibinfo{person}{Changhua Pei}, \bibinfo{person}{Qiang Su}, \bibinfo{person}{Jianfeng Wen}, \bibinfo{person}{Xueliang Wang}, \bibinfo{person}{Guanyu Mu}, \bibinfo{person}{Dong Zheng}, \bibinfo{person}{Peng Jiang}, {and} \bibinfo{person}{Kun Gai}.} \bibinfo{year}{2022}\natexlab{}.
\newblock \showarticletitle{Deconfounding Duration Bias in Watch-time Prediction for Video Recommendation}. In \bibinfo{booktitle}{\emph{Proceedings of the 28th ACM SIGKDD Conference on Knowledge Discovery and Data Mining}} (Washington DC, USA) \emph{(\bibinfo{series}{KDD '22})}. \bibinfo{publisher}{Association for Computing Machinery}, \bibinfo{address}{New York, NY, USA}, \bibinfo{pages}{4472–4481}.
\newblock
\showISBNx{9781450393850}
\urldef\tempurl%
\url{https://doi.org/10.1145/3534678.3539092}
\showDOI{\tempurl}


\bibitem[Zhang et~al\mbox{.}(2024b)]%
        {gen_agent}
\bibfield{author}{\bibinfo{person}{An Zhang}, \bibinfo{person}{Yuxin Chen}, \bibinfo{person}{Leheng Sheng}, \bibinfo{person}{Xiang Wang}, {and} \bibinfo{person}{Tat-Seng Chua}.} \bibinfo{year}{2024}\natexlab{b}.
\newblock \showarticletitle{On generative agents in recommendation}. In \bibinfo{booktitle}{\emph{Proceedings of the 47th international ACM SIGIR conference on research and development in Information Retrieval}}. \bibinfo{pages}{1807--1817}.
\newblock


\bibitem[Zhang et~al\mbox{.}(2024c)]%
        {agentcf}
\bibfield{author}{\bibinfo{person}{Junjie Zhang}, \bibinfo{person}{Yupeng Hou}, \bibinfo{person}{Ruobing Xie}, \bibinfo{person}{Wenqi Sun}, \bibinfo{person}{Julian McAuley}, \bibinfo{person}{Wayne~Xin Zhao}, \bibinfo{person}{Leyu Lin}, {and} \bibinfo{person}{Ji-Rong Wen}.} \bibinfo{year}{2024}\natexlab{c}.
\newblock \showarticletitle{Agentcf: Collaborative learning with autonomous language agents for recommender systems}. In \bibinfo{booktitle}{\emph{Proceedings of the ACM on Web Conference 2024}}. \bibinfo{pages}{3679--3689}.
\newblock


\bibitem[Zhang et~al\mbox{.}(2023b)]%
        {instructrec}
\bibfield{author}{\bibinfo{person}{Junjie Zhang}, \bibinfo{person}{Ruobing Xie}, \bibinfo{person}{Yupeng Hou}, \bibinfo{person}{Wayne~Xin Zhao}, \bibinfo{person}{Leyu Lin}, {and} \bibinfo{person}{Ji{-}Rong Wen}.} \bibinfo{year}{2023}\natexlab{b}.
\newblock \showarticletitle{Recommendation as Instruction Following: {A} Large Language Model Empowered Recommendation Approach}.
\newblock \bibinfo{journal}{\emph{CoRR}}  \bibinfo{volume}{abs/2305.07001} (\bibinfo{year}{2023}).
\newblock
\urldef\tempurl%
\url{https://doi.org/10.48550/ARXIV.2305.07001}
\showDOI{\tempurl}


\bibitem[Zhang et~al\mbox{.}(2024a)]%
        {binllm}
\bibfield{author}{\bibinfo{person}{Yang Zhang}, \bibinfo{person}{Keqin Bao}, \bibinfo{person}{Ming Yan}, \bibinfo{person}{Wenjie Wang}, \bibinfo{person}{Fuli Feng}, {and} \bibinfo{person}{Xiangnan He}.} \bibinfo{year}{2024}\natexlab{a}.
\newblock \showarticletitle{Text-like Encoding of Collaborative Information in Large Language Models for Recommendation}. In \bibinfo{booktitle}{\emph{Proceedings of the 62nd Annual Meeting of the Association for Computational Linguistics (Volume 1: Long Papers)}}. \bibinfo{publisher}{Association for Computational Linguistics}, \bibinfo{address}{Bangkok, Thailand}, \bibinfo{pages}{9181--9191}.
\newblock
\urldef\tempurl%
\url{https://doi.org/10.18653/v1/2024.acl-long.497}
\showDOI{\tempurl}


\bibitem[Zhang et~al\mbox{.}(2021)]%
        {pda}
\bibfield{author}{\bibinfo{person}{Yang Zhang}, \bibinfo{person}{Fuli Feng}, \bibinfo{person}{Xiangnan He}, \bibinfo{person}{Tianxin Wei}, \bibinfo{person}{Chonggang Song}, \bibinfo{person}{Guohui Ling}, {and} \bibinfo{person}{Yongdong Zhang}.} \bibinfo{year}{2021}\natexlab{}.
\newblock \showarticletitle{Causal Intervention for Leveraging Popularity Bias in Recommendation}. In \bibinfo{booktitle}{\emph{Proceedings of the 44th International ACM SIGIR Conference on Research and Development in Information Retrieval}} (Virtual Event, Canada) \emph{(\bibinfo{series}{SIGIR '21})}. \bibinfo{publisher}{Association for Computing Machinery}, \bibinfo{address}{New York, NY, USA}, \bibinfo{pages}{11–20}.
\newblock
\showISBNx{9781450380379}
\urldef\tempurl%
\url{https://doi.org/10.1145/3404835.3462875}
\showDOI{\tempurl}


\bibitem[Zhang et~al\mbox{.}(2023a)]%
        {collm}
\bibfield{author}{\bibinfo{person}{Yang Zhang}, \bibinfo{person}{Fuli Feng}, \bibinfo{person}{Jizhi Zhang}, \bibinfo{person}{Keqin Bao}, \bibinfo{person}{Qifan Wang}, {and} \bibinfo{person}{Xiangnan He}.} \bibinfo{year}{2023}\natexlab{a}.
\newblock \showarticletitle{Collm: Integrating collaborative embeddings into large language models for recommendation}.
\newblock \bibinfo{journal}{\emph{arXiv preprint arXiv:2310.19488}} (\bibinfo{year}{2023}).
\newblock


\bibitem[Zhang and Yang(2021)]%
        {multitask}
\bibfield{author}{\bibinfo{person}{Yu Zhang} {and} \bibinfo{person}{Qiang Yang}.} \bibinfo{year}{2021}\natexlab{}.
\newblock \showarticletitle{A survey on multi-task learning}.
\newblock \bibinfo{journal}{\emph{IEEE transactions on knowledge and data engineering}} \bibinfo{volume}{34}, \bibinfo{number}{12} (\bibinfo{year}{2021}), \bibinfo{pages}{5586--5609}.
\newblock


\bibitem[Zheng et~al\mbox{.}(2024b)]%
        {lcrec}
\bibfield{author}{\bibinfo{person}{Bowen Zheng}, \bibinfo{person}{Yupeng Hou}, \bibinfo{person}{Hongyu Lu}, \bibinfo{person}{Yu Chen}, \bibinfo{person}{Wayne~Xin Zhao}, \bibinfo{person}{Ming Chen}, {and} \bibinfo{person}{Ji-Rong Wen}.} \bibinfo{year}{2024}\natexlab{b}.
\newblock \showarticletitle{Adapting large language models by integrating collaborative semantics for recommendation}. In \bibinfo{booktitle}{\emph{2024 IEEE 40th International Conference on Data Engineering (ICDE)}}. IEEE, \bibinfo{pages}{1435--1448}.
\newblock


\bibitem[Zheng et~al\mbox{.}(2024a)]%
        {trsr}
\bibfield{author}{\bibinfo{person}{Zhi Zheng}, \bibinfo{person}{Wenshuo Chao}, \bibinfo{person}{Zhaopeng Qiu}, \bibinfo{person}{Hengshu Zhu}, {and} \bibinfo{person}{Hui Xiong}.} \bibinfo{year}{2024}\natexlab{a}.
\newblock \showarticletitle{Harnessing large language models for text-rich sequential recommendation}. In \bibinfo{booktitle}{\emph{Proceedings of the ACM on Web Conference 2024}}. \bibinfo{pages}{3207--3216}.
\newblock


\bibitem[Zhu et~al\mbox{.}(2024)]%
        {cllm4rec}
\bibfield{author}{\bibinfo{person}{Yaochen Zhu}, \bibinfo{person}{Liang Wu}, \bibinfo{person}{Qi Guo}, \bibinfo{person}{Liangjie Hong}, {and} \bibinfo{person}{Jundong Li}.} \bibinfo{year}{2024}\natexlab{}.
\newblock \showarticletitle{Collaborative Large Language Model for Recommender Systems}. In \bibinfo{booktitle}{\emph{Proceedings of the ACM Web Conference 2024}} (Singapore, Singapore) \emph{(\bibinfo{series}{WWW '24})}. \bibinfo{pages}{3162–3172}.
\newblock
\showISBNx{9798400701719}


\end{thebibliography}

\appendix

\end{document}